\documentclass{jfm}
\usepackage{amssymb,amsmath}

\usepackage{array}
\usepackage{latexsym}
\usepackage{bezier}
\usepackage{psfrag,graphicx}
\usepackage{bm}
\usepackage{float}
\usepackage{multirow}
\usepackage{mathrsfs}
\usepackage{color}
\usepackage[hidelinks]{hyperref}
\usepackage{xcolor}
\usepackage{marginnote}

\newcommand{\rmi}{\mathrm{i}}
\newcommand{\rme}{\mathrm{e}}

\newcommand{\vv}{\mathbf{v}}

\newcommand{\vPhi}{\boldsymbol\Phi}

\shortauthor{R. Ayats et al.}
\shorttitle{Parallelogram domain computation}

\title[Fully nonlinear mode competitions in magnetised Taylor-Couette
  flow]{Fully nonlinear mode competitions in magnetised Taylor-Couette
  flow} \author[R. Ayats, K. Deguchi, F. Mellibovsky and
  A. Meseguer]{R. Ayats$^1$, K. Deguchi$^2$, F. Mellibovsky$^1$ and
  A. Meseguer$^1$ \thanks{Email address for correspondence:
    roger.ayats@upc.edu, kengo.deguchi@monash.edu,
    fernando.mellibovsky@upc.edu, alvaro.meseguer@upc.edu}}

\affiliation{ \aff{1} Departament de F{\'\i}sica, Universitat
  Polit\`ecnica de Catalunya, 08034, Barcelona, Spain \aff{2} School
  of Mathematics, Monash University, VIC 3800, Australia }

\begin{document}

\maketitle

\begin{abstract}
We study the nonlinear mode competition of various spiral
instabilities in magnetised Taylor-Couette flow. The resulting
finite-amplitude mixed-mode solution branches are tracked using the
annular-parallelogram periodic domain approach developed by
\citet{DeAlt2013}. Mode competition phenomena are studied in both the
anti-cyclonic and cyclonic Rayleigh-stable regimes.  In the
anti-cyclonic sub-rotation regime, with the inner cylinder rotating
faster than the outer, \citet*{HTR10} found competing axisymmetric and
non-axisymmetric magneto-rotational linearly unstable modes within the
parameter range where experimental investigation is feasible.  Here we
confirm the existence of mode competition and compute the nonlinear
mixed-mode solutions that result from it.  In the cyclonic
super-rotating regime, with the inner cylinder rotating slower than
the outer, \citet{Deguchi_PRE17} recently found a non-axisymmetric
purely hydrodynamic linear instability that coexists with the
non-axisymmetric magneto-rotational instability discovered a little
earlier by \citet*{RSGS16}.
We show that nonlinear interactions of these instabilities give rise
to rich pattern-formation phenomena leading to drastic angular momentum transport
enhancement/reduction.
\end{abstract}

\section{Introduction}
\begin{figure}
\centering
\includegraphics[scale=0.4]{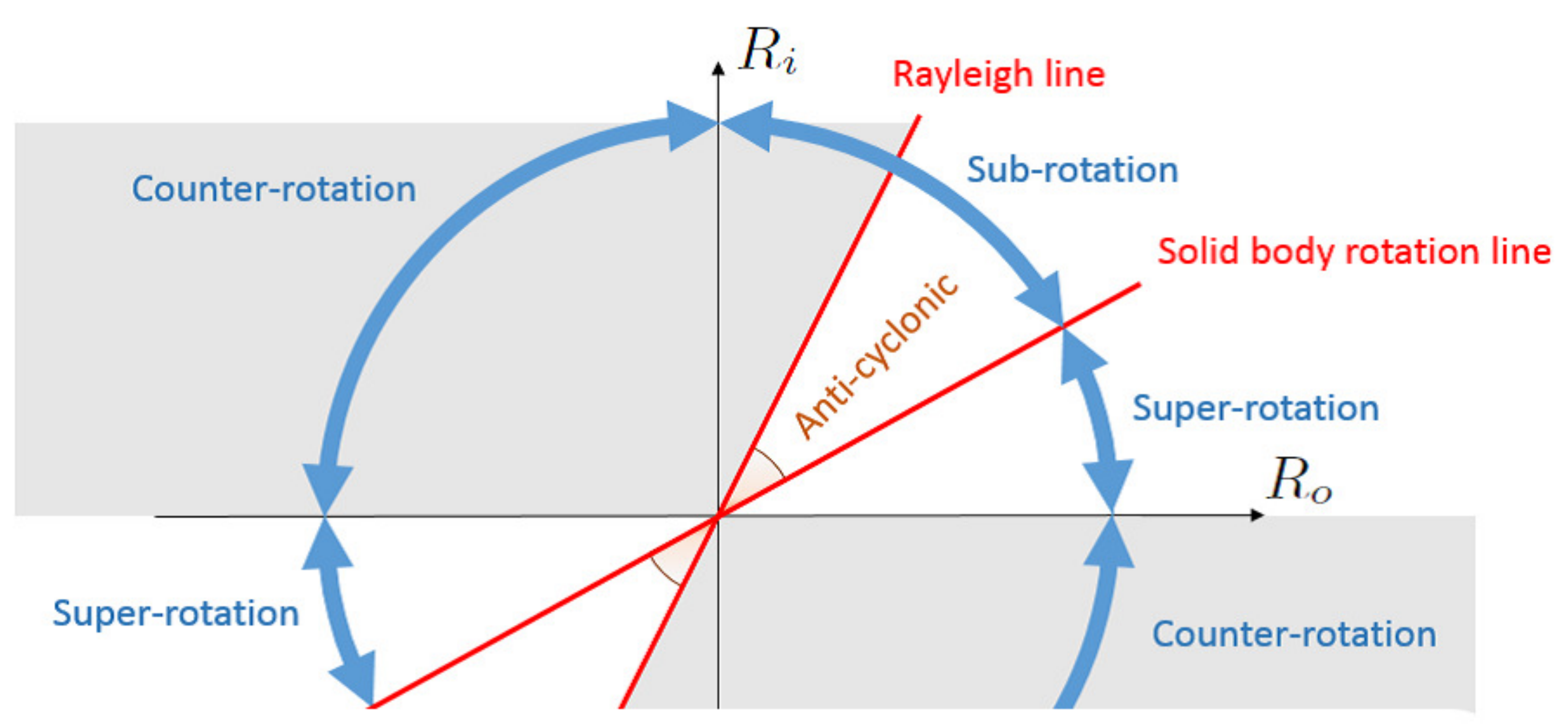} 
\caption{$R_i$--$R_o$ parameter space representation of Taylor-Couette
  flow. The plane is divided into the counter-rotation and the
  co-rotation regimes. This latter is further subdivided into
  sub-rotation and super-rotation by the solid-body rotation line
  defined by equal angular speed of both cylinders. The shaded regions
  denote inviscid instability of circular Couette
  flow following the Rayleigh criterion. The portion of the
  subrotation regime comprised between the Rayleigh and solid-body
  rotation lines goes by the name of anti-cyclonic regime (also
    called \textit{Quasi-Keplerian}), while the rest of the plane is
  called cyclonic.
}
\label{fig:param_plane}
\end{figure}

The objective of this study is the nonlinear interactions between
various instability modes occurring in magnetised Taylor-Couette flow,
i.e. the fluid flow between independently rotating concentric
cylinders.  The purely hydrodynamic Taylor-Couette flow, in the
absence of magnetic field, has long served as a theoretical, numerical
and experimental test bench for the study of centrifugal and shear
instability mechanisms. Keeping the outer cylinder stationary,
\citet{Ta1923} observed that the flow is destabilised by purely
hydrodynamic axisymmetric perturbations at a certain critical speed of
the inner cylinder.  The balance between rotational and shear effects
can be modified by further introducing a rotation of the outer
cylinder. The independent variation of the inner and outer cylinder
speeds results in a rich diversity of secondary nonlinear flow
patterns, as reported by \citet{ALS86}. The stability and nonlinear
states of Taylor-Couette flow are commonly studied in the $R_i$--$R_o$
parameter space schematically depicted in
figure~\ref{fig:param_plane}, where $R_i$ and $R_o$ are the Reynolds
numbers associated with the inner and outer cylinder speeds,
respectively. In accordance with the symmetries of the problem, the
$R_i$-$R_o$ parameter plane is invariant to $\pi$-rotation about the
origin, such that only the upper half plane needs to be explored. The
right/left half of the semi-plane corresponds to cylinders rotating in
the same/opposite direction (i.e. co-rotation/counter-rotation).  The
first quadrant (co-rotation regime) is divided into sub-rotation and
super-rotation by the solid-body rotation line (equal angular speed of
the cylinders), depending on whether the outer cylinder rotates slower
or faster than the inner. For any given speed of the outer cylinder,
Rayleigh's inviscid stability criterion establishes that circular
Couette flow remains centrifugally stable to infinitesimal
axisymmetric perturbations as long as the inner cylinder is steady or
in co-rotation up to a certain speed, delimited by the so-called
Rayleigh line (the wedge-shaped white region in
figure~\ref{fig:param_plane} comprised between the Rayleigh line and
the horizontal $R_i=0$ line). Taking viscous effects and
non-aximsymmetric perturbations into account affects the stability
boundaries, but it is widely accepted that the Rayleigh line acts as a
fairly approximate threshold below which circular
  Couette flow remains the only stable state, given that no
  experimental or numerical evidence of nonlinear flow states has been
  found to date \citep[see][]{Ji2006,Edlund2014,Lopez2017}.  Note
  however that no first-principle theory has been advanced so far to
  support the nonlinear hydrodynamic stability in the quasi-Keplerian
  flow regime \citep[see][for a summary on the
      matter]{Balbus2017}.

The Rayleigh-stable region is further subdivided into the
\textit{anti-cyclonic} and \textit{super-rotation cyclonic} regimes by
the solid body rotation line, corresponding to both cylinders rotating
at the same angular speed. Immediately to the right of the Rayleigh
line and all the way down to solid rotation, (co-rotation and
sub-rotation) is said to be \textit{anti-cyclonic} and laminar Couette
flow is allegedly linearly stable. This region is of utmost
astrophysical interest, since Keplerian rotational flow, a vastly used
model for accretion disks, is precisely anti-cyclonic. The rate at
which angular momentum is radially transported in astrophysical
accretion disks observation requires the flow to be turbulent. This
has motivated a wealth of studies (see the review paper
\citet{RuGeHoSchuSte18}) exploring magneto-rotational instabilities
(henceforth referred to as {\sc mri}) as a possible alternative source
for turbulence in a flow that appears otherwise
to always revert back to laminar in the absence of magnetic fields. The
pioneering works by \citet{Vel59} and \citet{Cha60} showed that a
uniform external magnetic field in the axial direction indeed
destabilises the anti-cyclonic regime, while the importance of the
instability in the astrophysical context was noted for the first time
by \citet{BaHa91}. This type of {\sc mri} is nowadays called the
\textit{standard} type of {\sc mri}, {\sc smri} for short.

A different approach was nevertheless taken in the first experimental
observation of the {\sc mri} \citep{SGGRSSH06}.  The {\sc smri} is
actually very difficult to reproduce in liquid metal experiments where
the magnetic Prandtl number $P_m$ is very small, as
  magnetic induction is essential in this case. As
shown by \citet{GoJi02}, the critical Reynolds number of the {\sc
  smri} is inversely proportional to the magnetic Prandtl number $P_m$
for small $P_m$, meaning that the cylinders must rotate at an
extremely fast rate to trigger the {\sc smri} in the experimental
apparatus.  The crux in reproducing a {\sc mri} at relatively small
Reynolds numbers was the numerical finding by \citet{HoRu05} that,
when both azimuthal and axial external magnetic fields are applied
simultaneously, the critical Reynolds number saturates at a finite
value even in the inductionless limit of
$P_m\rightarrow 0$. Soon after the discovery of this \textit{helical}
{\sc mri} (henceforth referred to as {\sc hmri}), growth of
axisymmetric perturbations was confirmed in the series of {\sc
  promise} experiments \citep{SGGRSSH06,SGGRSH07,RHSGGR06}. Later on,
\citet{HTR10} found that non-axisymmetric modes arise
  instead when purely azimuthal magnetic fields are considered in the
  sub-rotation regime just below the Rayleigh line. They further
  showed that these so-called azimuthal magneto-rotational instability
  ({\sc amri}) modes 
  persist when a small axial magnetic field is added to the
  predominantly azimuthal field, thus implying that they could
  potentially interact with the axisymmetric {\sc hmri}
  mode. In fact, when
the strength of the azimuthal and axial external magnetic fields are
suitably adjusted, the critical Reynolds numbers for the axisymmetric
and non-axisymmetric modes become comparable. The competition of these
modes may yield rich nonlinear flow patterns at this particular {\sc
  hmri} regime.  This nonlinear mode interaction is the first subject
we will tackle in this study. 

In the early years of the pattern-formation theoretical studies in
purely hydrodynamic Taylor-Couette flow, weakly nonlinear analysis was
employed to investigate mode interactions among multiple linear
instability modes near criticality
\citep[e.g.][]{DaDiSt68,Io86,GoStSc88,ChoIoo94}.  The simplest mode
interaction occurs between two identical but mutually-symmetric, with
respect to an axial reflection, spiral modes. In this case, the fully
nonlinear mixed-mode solution can be computed in numerical simulations
using a periodic axial-azimuthal orthogonal domain \citep{TESM89}.
However, when the interacting spirals are not mutually symmetric and
have a different absolute pitch, as are indeed the two mode
interactions studied here, the numerical computation of the fully
nonlinear mixed mode is no longer straightforward. The periodic
computational domain must fit an integer number of both constituent
modes in order to faithfully reproduce the mixed mode, which may lead
to unaffordably large domains.  This may be feasible in some occasions
\citep{PiLuHo06,AvMeMa2006,AlHo10}, but it is at the very least
inefficient from a computational point of view, if not altogether
prohibitive. A convenient methodology for the computation of general
mixed-mode states was provided by \citet{DeAlt2013}, who realised that
the infinite annulus might be subdivided into a regular tiling of a
suitable parallelogram-shaped periodic box that can be chosen
optimally small for any flow pattern arising from nonlinear
interaction of two modes, as we are considering here.
Extension of the nonlinear code to magnetised problems
  might prove highly valuable to the study of {\sc mri}, as nonlinear
  simulations in cylindrical/annular domains have only recently been
  undertaken \citep{Guseva2015,Guseva2017}.

In the second half of this paper we shall also study the nonlinear
mode competitions occurring in the other Rayleigh stable regime,
i.e. the super-rotation regime seen in figure 1.  Recently,
magnetohydrodynamic instabilities in this regime have attracted much
attention as they are thought to be relevant for turbulence generation
in a part of Sun's tachocline.  In this second subject we shall
investigate the nonlinear competition of two recently discovered
linear instability modes.

For the purely hydrodynamic problem, nonlaminar flow patterns in the
inviscidly stable super-rotation regime were first reported several
decades ago in Taylor-Couette experiments by \citet{Wendt33} and
\citet{Col65}, but at the time it was not clear whether the
instability was legitimate or an end-wall effect induced by the
cylinder lids. Advance in computational power eventually allowed to
numerically confirm the existence of subcritical spiral turbulence and
intermittency found experimentally
\citep{Van66,PriGreChaDauSaa02,HeAnHaPo89,BurCza12}
in the counter-rotation regime in the absence of end-wall effects
\citep{MeMeAvMa09_A,Dong2009}. Nonlinear coherent states have indeed
been followed into the super-rotation regime, crossing the $R_i=0$
boundary, as illustrated by the computation of the first rotating wave
in cyclonic super-rotation \citep*{DeMeMe14} and by direct numerical
simulation \citep*{OsVeLo16}.  All non-trivial flow patterns hitherto
observed in super-rotation are finite amplitude and highly nonlinear,
such that they by no means belie the widely assumed linear stability
of super-rotating hydrodynamic Taylor-Couette flow, in view of
countless numerical studies of the neutral curve (see the review
article by \citet{GroLoSu16}). The recent unexpected discovery by
\citet*{Deguchi_PRE17} of a linear instability in the super-rotation
regime came therefore as a big surprise. Considering non-axisymmetric
perturbations and a relatively long axial wavelength were key
ingredients to the finding.  This instability mode of a purely
hydrodynamic nature, hereafter called the {\sc d}17 mode, is the first
of the two modes we shall consider in our second mode competition
study.

The other mode at play inherently originates from the {\sc mri}
mechanism and is called the super-{\sc amri} \citep*{RSGS16,RSGS18},
where 
the prefix \emph{super} refers to the
  super-rotation regime. This mode belongs, along with the usual forms
  of {\sc hmri} and {\sc amri} for sub-rotation, to the class of
  inductionless {\sc mri}.  It has long been known
that {\sc mri} is not easily triggered in the super-rotation regime
for the axisymmetric case.  For ideal fluids, Velikhov's condition
states that the axial magnetic field cannot destabilise this regime
\citep{Vel59}, while according to Michael's condition \citep{Mi54}, an
azimuthal field can only be destabilising provided its modulus
increases outwards at a sufficiently fast rate.  Moreover, when the
azimuthal magnetic field is current-free, it can be formally shown
that axisymmetric {\sc mri} are impossible in spite of the diffusive
effect \citep{HeSo06}.  A breakthrough regarding instability in the
cyclonic super-rotation regime is due to \citet{StKi15}, who pointed
out that for sufficiently narrow gaps the non-axisymmetric instability could be continued into the
super-rotation regime using the so-called local approximation and the
inductionless limit ($P_m\rightarrow 0$).
However, the existence of the super-{\sc
  amri} was not conclusive at this stage given that a local
approximation does not always necessarily provide accurate insight
into the global problem. Soon after, conclusive numerical evidence of
the super-{\sc amri} was reported by \citet{RSGS16,RSGS18}, who
concluded that the destabilisation seems to occur for fairly arbitrary
magnetic field profiles as long as the flow is double-diffusive,
i.e. $P_m \neq 1$.

The paper is organised as follows. Section \S\ref{sec_formulation}
formulates the problem based on the inductionless limit of the
magneto-hydrodynamic equations. The section addresses in detail the
numerical discretization of the equations in annular-parallelogram
periodic domains and, in particular, describes the Newton solver for
the computation of nonlinear mixed-mode travelling waves using the
transformed coordinate system and a suitable co-moving reference
frame. Section \S\ref{sec_anticyclonic} is devoted to the
anti-cyclonic regime. The helical magnetic field is imposed to find
the nonlinear mixed-mode solutions that arise from the mode
competition advanced by \citet{HTR10}.  The first part of section
\S\ref{sec_counter_super} deals with the interaction between the
classical non-axisymmetric and the {\sc d}17 modes in purely
hydrodynamic counter-rotating Taylor-Couette flow. In the second half
of the section we shall see how an imposed azimuthal magnetic field
alters the nature of this interaction. Finally, in section
\S\ref{sec_conclusions}, we briefly summarise the results and present
concluding remarks.

\section{Formulation of the problem}
\label{sec_formulation}

Consider an electrically conducting fluid of density $\rho^*$,
kinematic viscosity $\nu^*$, and magnetic diffusivity $\eta^*$,
confined between two concentric cylinders of inner and outer radii
$r_i^*$ and $r_o^*$, independently rotating at angular speeds
$\Omega_i^*$ and $\Omega_o^*$, respectively.  In addition, the fluid
is subject to the action of a magnetic field of typical strength
$B_0^*$.  Throughout the paper we use the length $d^*=r_o^*-r_i^*$,
time $d^{*2}/\nu^*$, velocity $\nu^*/d^*$, and magnetic field
$\nu^*\sqrt{\rho^* \mu^*}/d^*$ scales for non-dimensionalisation,
where $\mu^*$ is the magnetic permeability.  As a consequence of using
the viscous time scale, the Reynolds numbers are absorbed into the
expression for the base state flow fields and disappear form the
non-dimensional equations for the perturbation.  The key parameters of
the flow are the radius ratio $\eta$, the inner $R_i$ and outer $R_o$
Reynolds numbers, along with the magnetic Prandtl number $P_m$, and
the Hartmann number $H$:
\begin{eqnarray}
\eta=\frac{r_i^*}{r_o^*},~~
R_i=\frac{\Omega_i^*r_i^*d^*}{\nu^*},~~
R_o=\frac{\Omega_o^*r_o^*d^*}{\nu^*},\qquad P_m=\frac{\nu^*}{\eta^*},~~
H=\frac{B_0^*d^*}{\sqrt{\rho^* \mu^* \eta^* \nu^*}}.\label{parameters}
\end{eqnarray}
The non-dimensional external magnetic field is proportional to
$P_m^{-1/2}H$. The reason for using $H$ is that we will consider the
so-called inductionless limit $P_m\rightarrow 0$ where $H$ is
typically fixed as a constant.

Non-dimensionalisation of the velocity
$\mathbf{v}=u\mathbf{e}_r+v\mathbf{e}_{\theta}+w\mathbf{e}_{z}$ and
magnetic
$\mathbf{B}=A\mathbf{e}_r+B\mathbf{e}_{\theta}+C\mathbf{e}_{z}$
fields, expressed here in cylindrical coordinates $(r,\theta,z)$,
yields the incompressible viscous-resistive {\sc mhd} equations
\begin{subequations}\label{govbody}
\begin{eqnarray}
\partial_t\mathbf{v}+(\mathbf{v}\cdot
\nabla)\mathbf{v}-(\mathbf{B}\cdot \nabla)\mathbf{B}=-\nabla
p+\nabla^2\mathbf{v},\\ \partial_t\mathbf{B}+(\mathbf{v}\cdot
\nabla)\mathbf{B}-(\mathbf{B}\cdot
\nabla)\mathbf{v}=P_m^{-1}\nabla^2\mathbf{B},\\ \nabla \cdot
\mathbf{v}=\nabla \cdot \mathbf{B}=0,
\end{eqnarray}
\end{subequations}
where $p$ is the total pressure and $t$ is time.  Equation
(\ref{govbody}a) expresses momentum conservation, while equation
(\ref{govbody}b) is the induction equation. Equations (\ref{govbody}c)
correspond to continuity and Gauss' law.  Along the cylinder walls at
radii
\begin{eqnarray}
r_i=\frac{\eta}{1-\eta},\qquad r_o=\frac{1}{1-\eta},
\label{ri_and_ro}
\end{eqnarray}
we assume no-slip and perfectly insulating boundary conditions. In our
formulation, the velocity and magnetic fields are decomposed into the
base and the perturbation flows following
\begin{subequations}
\begin{eqnarray}
  \mathbf{v}=v_b(r)\mathbf{e}_{\theta}+Gw_p(r)\mathbf{e}_{z}+
  \widetilde{\mathbf{v}}(r,\theta,z,t)\label{basev},\\
  \mathbf{B}=P_m^{-1/2}H\{B_b(r)\mathbf{e}_{\theta}+C_b(r)\mathbf{e}_{z}\}+
  \widetilde{\mathbf{B}}(r,\theta,z,t),
\end{eqnarray}
\end{subequations}
where the tilde denotes perturbation quantities. The pressure
perturbation is therefore written as $\widetilde{p}$.  Here
$v_b(r)=R_s r+R_pr^{-1}$ is the laminar \textit{Couette flow}
solution, with coefficients.
\begin{eqnarray}
R_s=\frac{R_o-\eta R_i}{1+\eta},\qquad R_p=\frac{\eta^{-1}R_i-R_o}{1+\eta}r_i^2,
\label{vb_couette}
\end{eqnarray}
where the subscripts denote the solid-body rotation ($s$) and the
potential ($p$) components of the flow.  External magnetic mechanisms
induce the base magnetic fields $B_b(r)$ and $C_b(r)$, which will be
duly introduced in (\ref{baseflowhtr10}) and (\ref{baseflowkengo}) for
the two types of predominantly azimuthal fields that will be
considered throughout the paper.

We will assume further that there is no axial net
mass flux.  This is accomplished by imposing an external instantaneously adjustable axial
pressure gradient that induces the well-known base \textit{annular Poiseuille flow} profile
\begin{eqnarray}
w_p(r)=(r^2-r_i^2)\ln r_o +(r_o^2-r^2)\ln r_i -(r_o^2-r_i^2)\ln r.
\end{eqnarray}
The product $G w_p$ in (\ref{basev}) represents the axial flow induced by the
external pressure gradient, whose strength is measured by the coefficient $G$.
That coefficient is a time-dependent additional unknown in the
constant mass flux problem.  For travelling wave states $G$ is merely
a constant. Moreover, it is easy to show that when the flow possesses
some symmetry in $z$, $G$ must vanish.

For liquid metals used in laboratory experiments $P_m$ is very small
$(10^{-5}\sim 10^{-7})$.  It is therefore reasonable to apply the
inductionless limit approximation $P_m\rightarrow 0$ to the governing
equations \citep[see][for example]{Davidson2017}.  The magnetic field
perturbation is rescaled as
$\widetilde{\mathbf{b}}=P_m^{-1/2}H^{-1}\widetilde{\mathbf{B}}$ and
the size of the variables
$\widetilde{\mathbf{v}},\widetilde{\mathbf{b}},\widetilde{p}$ and
$R_i, R_o, H$ are fixed as $O(P_m^0)$ quantities during the limiting
process. The resulting leading-order equations are
\begin{subequations}\label{lim}
\begin{eqnarray}
\left[
\begin{array}{c}
(\partial_t+r^{-1}v_b\partial_{\theta}+R_pw_p\partial_z)\widetilde{u}-2r^{-1}v_b\widetilde{v}\\
(\partial_t+r^{-1}v_b\partial_{\theta}+R_pw_p\partial_z)\widetilde{v}+r^{-1}(rv_b)' \widetilde{u}\\
(\partial_t+r^{-1}v_b\partial_{\theta}+R_pw_p\partial_z)\widetilde{w}  \end{array} \right]\hspace{60mm}\nonumber \\
-H^2\left[ \begin{array}{c}
(r^{-1}B_b\partial_{\theta}+C_b\partial_z) \widetilde{a}-2r^{-1}B_b\widetilde{b}\\
(r^{-1}B_b\partial_{\theta}+C_b\partial_z)\widetilde{b}+r^{-1}(rB_b)'\widetilde{a} \\
(r^{-1}B_b\partial_{\theta}+C_b\partial_z)\widetilde{c} \end{array} \right]
+(\widetilde{\mathbf{v}}\cdot \nabla)\widetilde{\mathbf{v}}=
-\nabla \widetilde{p}+\nabla^2\widetilde{\mathbf{v}},~~~~~~~~\label{limmo} \\
-\left[ \begin{array}{c}
(r^{-1}B_b\partial_{\theta}+C_b\partial_z)\widetilde{u}\\
(r^{-1}B_b\partial_{\theta}+C_b\partial_z)\widetilde{v}-r(r^{-1}B_b)'\widetilde{u}\\
(r^{-1}B_b\partial_{\theta}+C_b\partial_z)\widetilde{w}  \end{array} \right]
=\nabla^2\widetilde{\mathbf{b}},~~~~~~~~~~~~~~~~\label{limind}
\end{eqnarray}
\end{subequations}
along with the solenoidal conditions $\nabla \cdot
\widetilde{\mathbf{v}}=\nabla \cdot
\widetilde{\mathbf{b}}=0$. The time derivative drops
  out from the induction equations on account of applying the
  inductionless limit, and (\ref{limind}) becomes a
mere linear system linking the velocity and the magnetic field.  It
can thus be used, as will be shown shortly, to eliminate the magnetic
perturbation from the momentum equation (\ref{limmo}).

\subsection{Spectral discretisation on a parallelogram domain}
We shall be looking here for nonlinear travelling wave solutions of the above resulting equations.
The hydrodynamic and magnetic perturbation fields
$\widetilde{\mathbf{v}}$ and $\widetilde{\mathbf{b}}$ are both
solenoidal, so they admit a \textit{toroidal-poloidal} decomposition of
the form
\begin{subequations}
\begin{eqnarray}
\widetilde{\mathbf{v}}(r,\theta,z,t)=
\mathbf{e}_{\theta}\overline{v}(r)+\mathbf{e}_{z}\overline{w}(r)+
\nabla \times \nabla \times \{\mathbf{e}_r\phi(r,\theta,z,t) \}+ \nabla
\times \{\mathbf{e}_r\psi(r,\theta,z,t)
\},~~~\label{poltorpotphipsi}\\ \widetilde{\mathbf{b}}(r,\theta,z,t)=\nabla \times \nabla \times
\{\mathbf{e}_rf(r,\theta,z,t) \}+ \nabla \times
\{\mathbf{e}_rg(r,\theta,z,t) \},~~~~~~\label{poltorpotfg}
\end{eqnarray}
\end{subequations}
where $\overline{v}(r)$ and $\overline{w}(r)$ are the azimuthal and
axial components, respectively, of the mean velocity field. It can be
easily shown that no mean magnetic field can be generated in
the inductionless limit. The
poloidal and toroidal potentials $\phi, f$ and $\psi, g$ introduced in
(\ref{poltorpotphipsi}) and (\ref{poltorpotfg}) uniquely determine the physical
hydrodynamic and magnetic perturbation fields
$\widetilde{\mathbf{v}}$ and $\widetilde{\mathbf{b}}$, except for the obvious gauge freedom (addition of a constant).

\begin{figure}
\centering
\includegraphics[width=.5\linewidth,clip=]{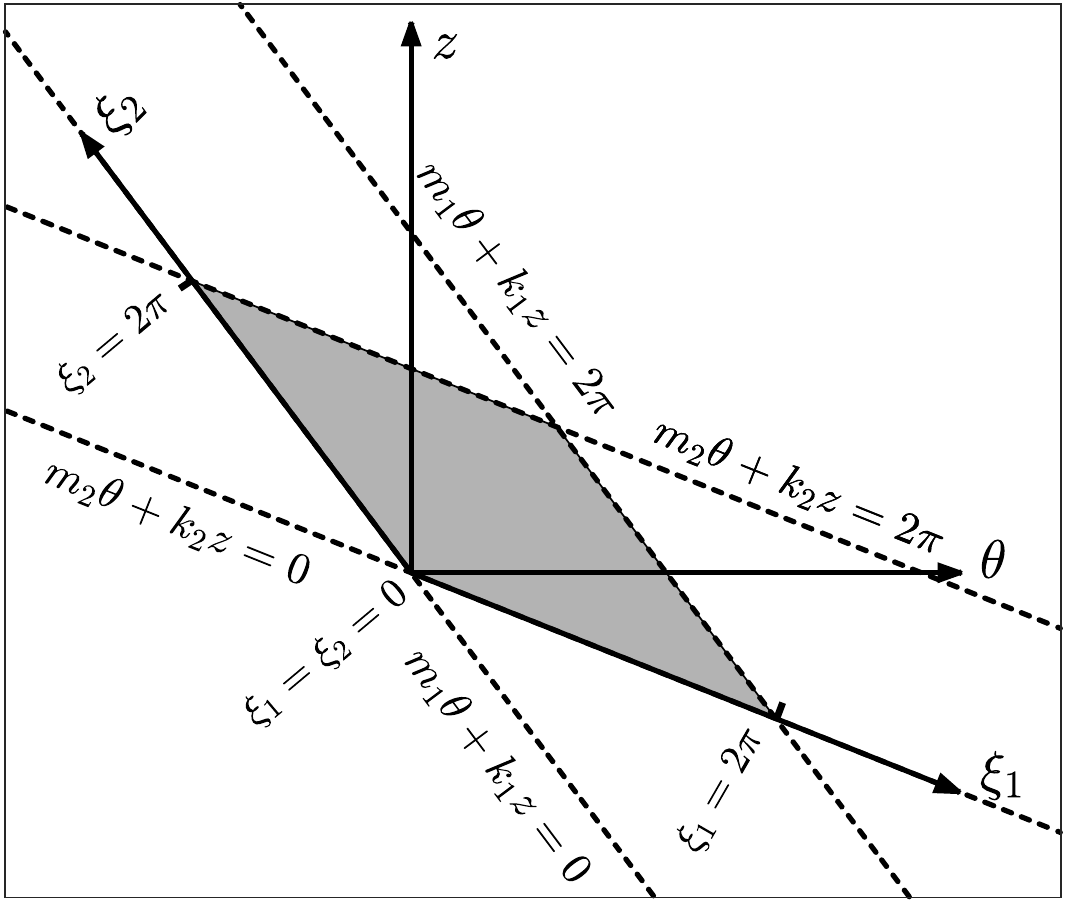} 
\caption{Sketch of the parallelogram domain introducing the new
  variables $(\xi_1,\xi_2)$ that replace the usual azimuthal and axial
  coordinates $(\theta,z)$.
}
\label{fig:parallel_scheme}
\end{figure}

The coherent flows addressed in this work are mixed modes resulting
from the nonlinear interaction of pairs of spiral waves propagating in
the $(\theta,z)$-plane. Following \citet{DeAlt2013}, we introduce
the two phase variables
\begin{eqnarray}
\xi_1=m_1\theta+k_1z-c_1t,\qquad 
\xi_2=m_2\theta+k_2z-c_2t,
\label{spiwavfronts}
\end{eqnarray}
describing the wavefronts of the two interacting spirals, which
propagate at speeds $c_1$ and $c_2$, and whose azimuthal and axial
wavenumbers are the integer $(m_1,m_2)$ and real $(k_1,k_2)$ constant
pairs, respectively.  Travelling mixed modes resulting from the
nonlinear interaction of spiral modes of the form given by
(\ref{spiwavfronts}) are naturally represented on doubly
$2\pi$-periodic parallelogram domains of the form
\begin{equation}
(r,\xi_1,\xi_2) \in \left[r_i,r_o\right] \times[0,2\pi]\times[0,2\pi],
\label{ann_paral_domain}
\end{equation}
unwrapped and outlined in figure~\ref{fig:parallel_scheme} for any given value of the radial coordinate.
Straightforward algebraic manipulation shows that any
function of $\xi_1, \xi_2$ can also be written in terms of
$\theta-c_{\theta}t$ and $z-c_zt$ with
\begin{eqnarray}
c_{\theta}=\frac{k_2c_1-k_1c_2}{m_2k_1-m_1k_2},\qquad
c_z=\frac{m_2c_1-m_1c_2}{m_2k_1-m_1k_2}.
\end{eqnarray}
The solutions sought are therefore travelling waves propagating both
azimuthally and axially with the phase speeds $c_{\theta}$ and $c_z$
just given, respectively.

The initial-boundary value problem (\ref{limmo})-(\ref{limind}) is
reformulated in the new phase variables assuming $2\pi$-periodicity of
the toroidal and poloidal potentials introduced in
(\ref{poltorpotphipsi})-(\ref{poltorpotfg})
\begin{equation}
  [\phi,\psi,f,g](r,\xi_1+2\pi,\xi_2) = [\phi,\psi,f,g](r,\xi_1,\xi_2+2\pi) = [\phi,\psi,f,g](r,\xi_1,\xi_2).
\label{perbc_pots}
\end{equation}
The potentials are then discretised using spectral Fourier
expansions of the form
\begin{subequations}\label{potentials}
\begin{eqnarray}
\phi(r,\xi_1,\xi_2)=\sum_{n_1,n_2}\widehat{\phi}_{n_1n_2}(r)\,\rme^{\rmi(n_1\xi_1+n_2\xi_2)},
\;
\psi(r,\xi_1,\xi_2)=\sum_{n_1,n_2}\widehat{\psi}_{n_1n_2}(r)\,\rme^{\rmi(n_1\xi_1+n_2\xi_2)},~~~~~~~\\
f(r,\xi_1,\xi_2)=\sum_{n_1,n_2}\widehat{f}_{n_1n_2}(r)\,\rme^{\rmi(n_1\xi_1+n_2\xi_2)},\;
g(r,\xi_1,\xi_2)=\sum_{n_1,n_2}\widehat{g}_{n_1n_2}(r)\,\rme^{\rmi(n_1\xi_1+n_2\xi_2)},~~~~~~~
\end{eqnarray}
\end{subequations}
where the Fourier radial functions $\widehat{\phi}_{n_1n_2}(r)$,
$\widehat{\psi}_{n_1n_2}(r)$, $\widehat{f}_{n_1n_2}(r)$, and
$\widehat{g}_{n_1n_2}(r)$ are identically zero for $n_1=n_2=0$. For $n_1 \neq 0$
or $n_2 \neq 0$, these radial functions are suitable expansions of
modified Chebyshev polynomials satisfying homogeneous no-slip boundary
conditions at the inner and outer cylinder walls
\begin{eqnarray}
\phi= \partial_r\phi =\psi=\overline{v}=\overline{w}=0.
\label{hbcpotentials}
\end{eqnarray}
The hydrodynamic radial functions are thus
\begin{subequations}
\begin{eqnarray}
\widehat{\phi}_{n_1n_2}(r)=\sum_l X^{(1)}_{ln_1n_2}(1-y^2)^2T_l(y),~~\label{specappx_phi}
\widehat{\psi}_{n_1n_2}(r)=\sum_l X^{(2)}_{ln_1n_2}(1-y^2)T_l(y),~~~~\\
\overline{v}(r)=\sum_l X^{(1)}_{l00}(1-y^2)T_l(y),~~
\overline{w}(r)=\sum_l X^{(2)}_{l00}(1-y^2)T_l(y),~~~~
\label{specappx_psi}
\end{eqnarray}
\end{subequations}
where $T_l(y)$ is the $l^{\mathrm{th}}$ Chebyshev polynomial and $y\equiv2(r-r_i)-1\in
[-1,1]$ is the rescaled radial coordinate. Similarly, the magnetic radial functions are expanded employing
Chebyshev polynomials modified to satisfy perfectly insulating
conditions
\begin{subequations}
\begin{eqnarray}
\widehat{f}_{n_1n_2}(r)=\sum_l
X^{(3)}_{ln_1n_2}\left\{(1-y^2)T_l(y)+\alpha_{ln_1n_2} +\beta_{ln_1n_2} y\right\},\label{specappx_f}\\
\widehat{g}_{n_1n_2}(r)=\gamma_{n_1n_2}(r)\widehat{f}_{n_1n_2}(r)+
\sum_l X^{(4)}_{ln_1n_2}(1-y^2)T_l(y),
\label{specappx_g}
\end{eqnarray}
\end{subequations}
where a detailed description of the coefficients $\alpha_{ln_1n_2}$,
$\beta_{ln_1n_2}$, and of the function $\gamma_{n_1n_2}(r)$ can be
found in Appendix A. 

For computational purposes, the Fourier-Chebyshev expansions
(\ref{specappx_phi}\,-\,\ref{specappx_g}) are truncated at $l=L$, $|n_1|=N_1$ and
$|n_2|=N_2$. 
After substituting the truncated expansions into
system (\ref{limmo}\,-\,\ref{limind}), 
these are then
evaluated at the Chebyshev nodes
\begin{equation}
y=\cos \left (\frac{l+1}{L+2}\pi \right ), \qquad (l=1,\dots,L).
\end{equation}
This procedure leads to a system of nonlinear algebraic equations of
the form
\begin{subequations}
\begin{equation}\label{eq:disceqH}
0=\mathbb{L}_1
\left [
\begin{array}{c}
X^{(1)}_{ln_1n_2}\\X^{(2)}_{ln_1n_2}
\end{array}
\right ]
+H^2\mathbb{L}_2
\left [
\begin{array}{c}
X^{(3)}_{ln_1n_2}\\X^{(4)}_{ln_1n_2}
\end{array}
\right ]
+
[X^{(1)}_{ln_1n_2}, X^{(2)}_{ln_1n_2}]\mathbb{N}
\left [
\begin{array}{c}
X^{(1)}_{ln_1n_2}\\X^{(2)}_{ln_1n_2}
\end{array}
\right ],
\end{equation}
\begin{equation}\label{eq:disceqM}
\mathbb{L}_3
\left [
\begin{array}{c}
X^{(1)}_{ln_1n_2}\\X^{(2)}_{ln_1n_2}
\end{array}
\right ]
=\mathbb{L}_4
\left [
\begin{array}{c}
X^{(3)}_{ln_1n_2}\\X^{(4)}_{ln_1n_2}
\end{array}
\right ].~~~
\end{equation}
\end{subequations}
Here $\mathbb{L}_1, \mathbb{L}_2, \mathbb{L}_3, \mathbb{L}_4$ are
matrices, and $\mathbb{N}$ is a third-order tensor, whose form is
unchanged from the purely hydrodynamic case.  Isolating the magnetic
unknowns by solving the linear system (\ref{eq:disceqM}) and
substituting into (\ref{eq:disceqH}) yields a nonlinear system of
equations for the hydrodynamic unknowns $X^{(1)}_{ln_1n_2}$ and
$X^{(2)}_{ln_1n_2}$
\begin{eqnarray}
0=(\mathbb{L}_1+H^2\mathbb{L}_2\mathbb{L}_4^{-1}\mathbb{L}_3)
\left [
\begin{array}{c}
X^{(1)}_{ln_1n_2}\\X^{(2)}_{ln_1n_2}
\end{array}
\right ]
+
[X^{(1)}_{ln_1n_2}, X^{(2)}_{ln_1n_2}]\mathbb{N}
\left [
\begin{array}{c}
X^{(1)}_{ln_1n_2}\\X^{(2)}_{ln_1n_2}
\end{array}
\right ].~~~
\label{nonlinsyst_tw}
\end{eqnarray}
Matrix $\mathbb{L}_1$ depends implicitly on the unknown speeds $c_1$
and $c_2$ appearing in (\ref{spiwavfronts}) and that correspond to the
co-moving reference frame in which the mixed mode remains a steady
solution. Since these two speeds are also unknown, two additional
\textit{phase-locking} conditions are required to lift the
rotational/travelling degeneracy of solutions from the system of
equations. Similarly, system (\ref{nonlinsyst_tw}) must also be
complemented with an additional constraint to allow determination of
the unknown axial pressure-gradient $G$ required to ensure the zero
mass-flux condition.  The nonlinear system of equations
(\ref{nonlinsyst_tw}), along with the aforementioned constraints, is
solved numerically using Newton's method. The
  hydrodynamic part of the code is identical to that used in Deguchi
  \& Altmeyer (2013), and more detailed documentation of the
  computational methodology can be found in \citet{DeNa2011}.

For the purely hydrodynamic problem, we have also computed and
continued in parameter space the bifurcating mixed modes using an
independent numerical formulation. This alternative methodology is
based on a solenoidal Petrov-Galerkin scheme described in
\citet{EPJSTMAMM07}, suitably adapted to the annular parallelogram
domain (\ref{ann_paral_domain}). In this formulation, the solenoidal
velocity perturbation $\widetilde{\mathbf{v}}$ is approximated by
means of a spectral expansion $\widetilde{\mathbf{v}}_{\rm s}$ of
order $N$ in $\xi_1=m_1\theta+k_1z$, order $L$ in
$\xi_2=m_2\theta+k_2z$, and order $M$ in $r$
\begin{equation}
 \widetilde{\mathbf{v}}_{\rm s}(r,\xi_1,\xi_2,t)=
\sum_{n_1,\,n_2,\,m}
  a_{n_1 n_2m}(t)\, \vPhi_{n_1 n_2 m}(r,\xi_1,\xi_2).
\label{specapx}
\end{equation}
The $\vPhi_{n_1 n_2 m}$ are {\em trial} bases of
solenoidal vector fields of the form
\begin{equation}
\vPhi_{n_1 n_2 m}(r,\xi_1,\xi_2)=
\rme^{\rmi(n_1\xi_1+n_2\xi_2)}
\vv_{n_1 n_2 m}(r),
\label{physbas}
\end{equation}
where the radial fields $\vv_{n_1 n_2 m}(r)$ are suitably constructed
to satisfy $\nabla \cdot \vPhi_{n_1 n_2 m} = 0$.  Since $\widetilde{\mathbf{v}}_{\rm s}$
represents the perturbation of the velocity field, it must therefore
vanish at the inner ($r=r_i$) and outer ($r=r_o$) walls of the
cylinders. Therefore, $\vv_{n_1 n_2 m}$ must also satisfy
homogeneous boundary conditions
\begin{equation}
\vv_{n_1 n_2 m}(r_i)=\vv_{n_1 n_2 m}(r_o)=\mathbf{0}.
\label{hbc_phi}
\end{equation}
These radial fields are built from suitable expansions of modified
Chebyshev polynomials. 
After introducing expansion  
\begin{equation}
  \widetilde{\mathbf{v}}_{\rm s}(r,\xi_1,\xi_2,t)=
\sum_{n_1,\,n_2,\,m}
a_{n_1 n_2 m}^{\rm TW}\rme^{\rmi n_1(\xi_1-c_1
  t)}\rme^{\rmi n_2(\xi_2-c_2 t)}
\vv_{n_1 n_2 m}(r)
\label{specapx_tw}
\end{equation}
into the hydrodynamic equations, the weak formulation described in
\citet{EPJSTMAMM07} leads to a system of nonlinear algebraic equations
for the unknown coefficients $a_{n_1 n_2 m}^{\rm TW}$, similar to
(\ref{nonlinsyst_tw}), to which the zero mass-flux constraint is also
imposed. The resulting system of equations were solved using a
\textit{matrix-free} Newton-Krylov method \citep{Kel03}. The converged
nonlinear solutions were then continued in parameter space using
pseudo-arclength continuation schemes \citep{Kuz04}. To avoid
cluttering the paper with unnecessary detail and because of the
intricacies that are inherent to the numerical approach undertaken, a
detailed description of the method will be published separately.

In the classic rectangular domain, the Petrov-Galerkin solenoidal
discretization has been successfully used in the numerical
approximation of transitional flows in cylindrical geometries
\citep{MeMe05} and in the computation of subcritical rotating waves in
annular domains \citep{DeMeMe14}.  In the latter study, the code was
cross-checked against the codes used in the aforementioned
\citet{DeNa2011} and \citet{DeAlt2013}.  In \S4, the favourable
comparison of the nonlinear results produced by the
annular-parallelogram extension of the two independent codes based on
completely different formulations serves as an unbeatable procedure
for code validation.  The results for the linear magnetic part of
(\ref{nonlinsyst_tw}) has instead been checked against the linear
results by \citet{HTR10} in the next section.

For a travelling wave solution, the absolute values of \textit{torque}
on the inner and outer cylinders are always equal and represent the
angular momentum transport. The torque on the inner cylinder can be
computed indistinctly as
\begin{eqnarray}
T \equiv   \{-r^3\partial_r(r^{-1}\overline{v})\}|_{r=r_i} = -\{r^3\partial_r(r^{-1}\overline{v})\}|_{r=r_o},\label{torquedef}
\end{eqnarray}
while the torque on the outer cylinder is $-T$ to keep the inner and
outer cylinder rotating at constant speeds.  We have characterized
  all Newton-converged nonlinear solutions throughout by their torque
  normalised by the corresponding base-flow torque
  $T_b=\{-r^3\partial_r(r^{-1}v_b)\}|_{r=r_i}=-\{r^3\partial_r(r^{-1}v_b)\}|_{r=r_o}$
\begin{eqnarray}
\tau=\frac{T}{T_b}=\left. \frac{\partial_r(r^{-1}\overline{v})}{\partial_r(r^{-1}v_b)}\right |_{r=r_i,r_o},
\end{eqnarray}
such that the normalized torque $\tau$ is unity for laminar cicular Couette flow.

\section{The anti-cyclonic regime}
\label{sec_anticyclonic}

Let us consider the normalised base magnetic fields
\begin{eqnarray}
B_b(r)=\frac{r_i}{r},\qquad C_b(r)=\delta
\label{baseflowhtr10}
\end{eqnarray}
to reproduce both the axisymmetric {\sc hmri} and
  non-axisymmetric {\sc amri} modes found in the
anti-cyclonic regime by \citet{HTR10}.  The constant $\delta$
represents the strength of the axial magnetic field relative to the
azimuthal field, which is induced by a current running through the
inner cylinder, parallel to its axis.
  
Following \citet{HTR10}, we fix the rotation ratio
to $\widehat{\mu}=\varOmega_o^*/\varOmega_i^*=R_o\eta/R_i=0.26$.  Note
that for the anti-cyclonic regime $\widehat{\mu}$
must remain in the interval $[0.25, 1]$, where the lower bound
corresponds to the Rayleigh line $\widehat{\mu}=\eta^2=0.25$, while
the upper bound embodies solid-body rotation.  The quasi-Keplerian
rotation regime frequently used in astrophysical studies on accretion
disks is charcterized by $\widehat{\mu}=\eta^{3/2}\approx 0.35$. This
rotation law results from applying Kepler's law to both the inner and
outer cylinder angular velocities, which results in a fair
approximation of a strictly Keplerian flow across the gap.  The choice
$\widehat{\mu}=0.26$, used in the experimental
  demonstration of {\sc amri} by
  \citet{Seilmayer2014} places the flow in the
anti-cyclonic regime but very close to the boundary set by the
Rayleigh line. \citet{Liu2006} used a locally periodic
  approach to show that there is a limiting value
  $\widehat{\mu}\approx 0.3$ above which {\sc hmri} halts, and the
  analysis was later extended by \citet{Kirillov2012} to {\sc amri}.
  To what extent this limit is actually relevant to fully cylindrical
  flows is however still under debate
  \citep[see][]{RH2007,Child2015}. The radius ratio
of the cylinders is set to $\eta=0.5$. For this particular value of
$\eta$, our definitions of $R_i$ and $H$ become identical to the
hydrodynamic Reynolds number and the Hartmann number, respectively,
used by \citet{HTR10}. Most importantly, the parameter range studied
there is feasible in the {\sc
    promise} experiments, where both axisymmetric
  \citep{SGGRSSH06,SGGRSH07,RHSGGR06} and non-axisymmetric
  \citep{Seilmayer2014} modes were actually
realised. Travelling waves similar to those predicted
  in the numerical studies were indeed observed. These waves originate from absolute instability (even global), rather than mere convective,
  as shown by the comprehensive experimental study on {\sc hmri}
  by \citet{SGGHPRS09}.

\begin{figure}
\begin{center}
\begin{tabular}{cc}
%  \raisebox{0.42\linewidth}{(a)} & \includegraphics[width=.65\linewidth,clip=]{fig3a.eps}\\
%  \raisebox{0.42\linewidth}{(b)} &  \includegraphics[width=.65\linewidth,clip=]{fig3b.eps}
  \raisebox{0.42\linewidth}{(a)} & \includegraphics[width=.65\linewidth,clip=]{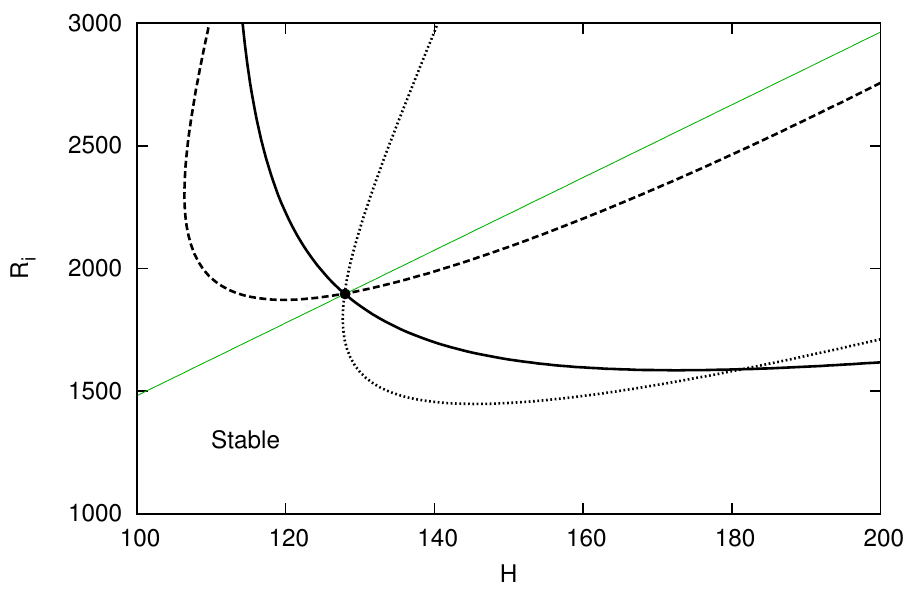}\\
  \raisebox{0.42\linewidth}{(b)} &  \includegraphics[width=.65\linewidth,clip=]{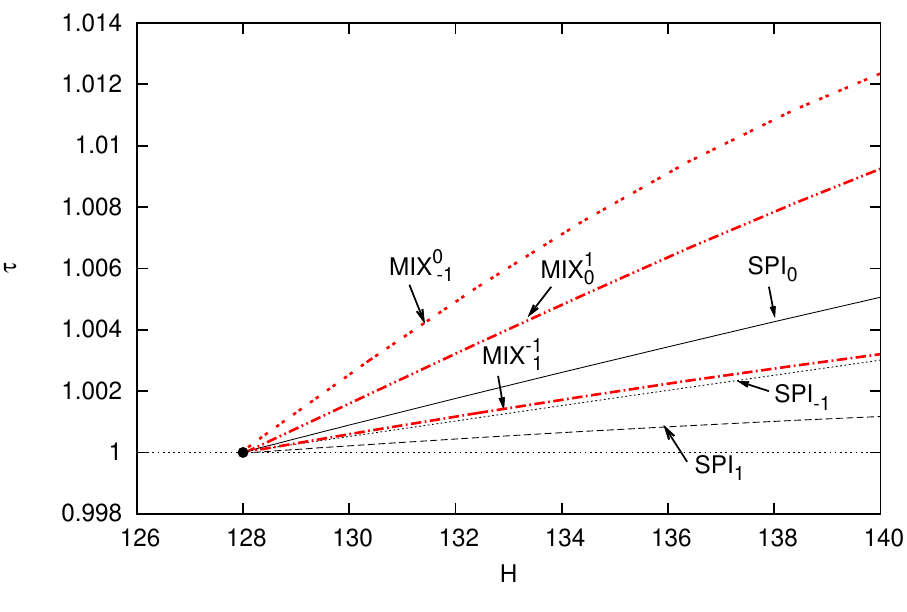}
\end{tabular}
\end{center}
\caption{Linear stability analysis and continuation of bifurcated
  nonlinear solution branches in the anti-cyclonic regime for
  $(\delta,\eta)=(0.0413,0.5)$. (a) Neutral stability curves along the
  line $R_o=0.26R_i/\eta$ (black curves) for modes $m=0$ (solid) and
  $m=\pm1$ (dashed for +1, dotted for -1).
  Wavenumber $k$ is the one that maximises growth rate.  The black
  circle indicates the triple critical point at
  $(H,R_i)=(128,1896)$. (b) Bifurcation diagram along
  $R_i=(1896/128)H$ (green line in pannel a). The black circle
  corresponds again to the triple-critical point, whence three spiral
  ($\textsc{spi}_0$, $\textsc{spi}_{\pm1}$; solid, dashed and dotted
  black lines) and three mixed ($\textsc{mix}_1^{-1}$,
  $\textsc{mix}_1^0$, $\textsc{mix}_0^{-1}$; solid, dashed and dotted
  red lines) modes are issued.}
\label{fig:nc_bifdiag}
\end{figure}

Having fixed $\widehat{\mu}$ and $\eta$, we have performed a linear
stability analysis of the base flow by exploring the eigenspectrum of
the linearized hydromagnetic equations for combinations of $R_i$, $H$,
$\delta$, and azimuthal-axial pairs $(m,k)$ of the associated spiral
eigenfunctions. We started by reproducing the neutral curves in the
$H$--$R_i$ plane for $\delta=0, 0.02, 0.03, 0.04, 0.05$, and for the
optimal axial wavenumber $k>0$ that maximizes the growth rate. For
$\delta=0$, the instability originates from the symmetric spirals with
opposite tilt $(m=\pm1)$.  The primal effect of finite $\delta$ is the
breaking of that reflection symmetry.  Moreover, the axisymmetric mode
$(m=0)$ emerges and for sufficiently large $\delta\approx 0.05$ it
dominates over the non-axisymmetric modes.  The
  non-axisymmetric modes are of {\sc amri} origin, while the
  axisymmetric mode only becomes dominant for distinctly helical
  fields, which leaves a finite range of $\delta$ where all three
  modes compete. The neutral curves we have computed
are in perfect agreement with figure 3 of \citet{HTR10}, where it was
already pointed out that, for $\delta \approx 0.04$, the critical
Reynolds numbers of all three modes become comparable.  Here we have
identified that at $\delta=0.0413$ there is a point where all three
modes become neutral simultaneously, as clearly shown in
figure\,\ref{fig:nc_bifdiag}.

For the sake of clarity, we shall focus on the computation and
continuation of nonlinear solutions along the straight line across
parameter space $R_i=(1896/128)H$ (green solid line in
figure~\ref{fig:nc_bifdiag}a) that passes through the triple critical
point at $(H,R_i)=(128, 1896)$.  Figure~\ref{fig:nc_bifdiag}b depicts
the bifurcation diagram corresponding to the $6$ different nonlinear
solution branches, as characterized by torque as a function of the
Hartmann number. At the tricritical point, the three eigenmodes,
$(m,k)=(0,5.672)$, $(m,k)=(1, 4.672)$ and $(m,k)= (-1,2.818)$, become
neutral simultaneously. As anticipated by weakly nonlinear analysis,
bifurcation of various mixed modes is therefore expected. The Newton
method described in \S\ref{sec_formulation} does indeed converge to
one or another of the nonlinear pure- or mixed-mode solutions when a
suitably weighted superposition of the three neutral eigenmodes is
taken as an initial guess. Solutions have initially been computed in
this way in the close neighbourhood of the tricritical point and then
continued as a function of $H$ using either natural or
pseudo-arclength continuation algorithms.  Three of the solution
branches, converged from single-mode initial guesses fed into the
Newton method, correspond to helically-invariant travelling spiral
waves (black curves in figure~\ref{fig:nc_bifdiag}b).  These solutions
we have dubbed as {\sc spi}$_{m}$, with the subscript $m$ (solid black
line for $m=0$, dashed for $m=1$, dotted for $m=-1$) denoting the
azimuthal wavenumber of the mode (see table~\ref{tab_samples}). All
three branches bifurcate supercritically from the base laminar
flow. Since these solutions can be computed in the usual rectangular
domain using regular coordinates $(\theta, z)$, we omit a detailed
analysis.
\begin{table}
  \begin{center}
%   \begin{minipage}{\hsize}
    \begin{tabular}{cccccc}
     Figure No. & Abbreviation & Solution type & ($m_1,k_1$) & ($m_2,k_2$)  \\
     \hline
     \ref{fig:nc_bifdiag}b & $\textsc{spi}_0$ & Spiral & (0,5.672) & {\sc n/a}  \\
     & $\textsc{spi}_1$ & Spiral & (1,4.672) & {\sc n/a} \\
     & $\textsc{spi}_{-1}$ & Spiral & (-1,2.818) & {\sc n/a} \\
     & $\textsc{mix}^{-1}_0$ & Mixed mode & (0,5.672) &  (-1,2.818)  \\
     & $\textsc{mix}^0_1$ & Mixed mode & (1,4.672) & (0,5.672) \\
     & $\textsc{mix}^{-1}_{1}$ & Mixed mode & (1,4.672)&  (-1,2.818) \\
     \hline
     \ref{fig:SuperbifSPa}a & $\textsc{spi}$ & Spiral & (1,40.6) & {\sc n/a}  \\
     & $\textsc{spi}_{\text{D}17}$ & Spiral & (1,1.002) & {\sc n/a} \\    
     & $\textsc{rib}$ & Ribbon & (1,40.6)  & (-1,40.6)   \\
     & $\textsc{rib}_{\text{D}17}$ & Ribbon & (1,1.002) & (-1,1.002)  \\     
     & $\textsc{mix}_{+}$ & Mixed mode & (1,40.6)  & (1,1.002)  \\
     & $\textsc{mix}_{-}$ & Mixed mode & (1,40.6) & (-1,1.002)  \\  
     \hline
     \ref{fig:SuperbifSPa}b & $\textsc{spi}_{\text{D}17}$ & Spiral & (1,0.616) & {\sc n/a}  \\
     & $\textsc{spi}_{\text{MRI}}$ & Spiral & (1,1.984) & {\sc n/a} \\    
     & $\textsc{rib}_{\text{D}17}$ & Ribbon & (1,0.616)  & (-1,0.616)   \\
     & $\textsc{rib}_{\text{MRI}}$ & Ribbon & (1,1.984) & (-1,1.984)  \\     
     & $\textsc{mix}_{+}$ & Mixed mode & (1,1.984) & (1,0.616)  \\
     & $\textsc{mix}_{-}$ & Mixed mode & (1,1.984) & (-1,0.616)  \\  
     \hline
    \end{tabular}
  \end{center}
\caption{Abbreviations used to describe the various nonlinear solution
  branches. Note that \textsc{spi}$_0$ is a zero-pitch spiral, and
  therefore a toroidal-vortex-pair solution.}
\label{tab_samples}
\end{table}

Suitable proportions of the weights applied to the critical eigenmodes
in generating the initial seeds for the Newton method have allowed
computation of all three possible mixed-mode nonlinear solution
branches (red curves in figure~\ref{fig:nc_bifdiag}b). These mixed
modes have been labeled as $\textsc{mix}_{m_1}^{m_2}$, with the sub-
and super- scripts representing the azimuthal wavenumber of the two
interacting modes, and duly reported
in table~\ref{tab_samples}.  As mentioned earlier, these mixed modes
can only be identified using an appropriate annular-parallelogram
domain, as the superposition of the two modes does not fit any
rectangular domain of affordable size.

\begin{figure}
\centering
\includegraphics[scale=1.2]{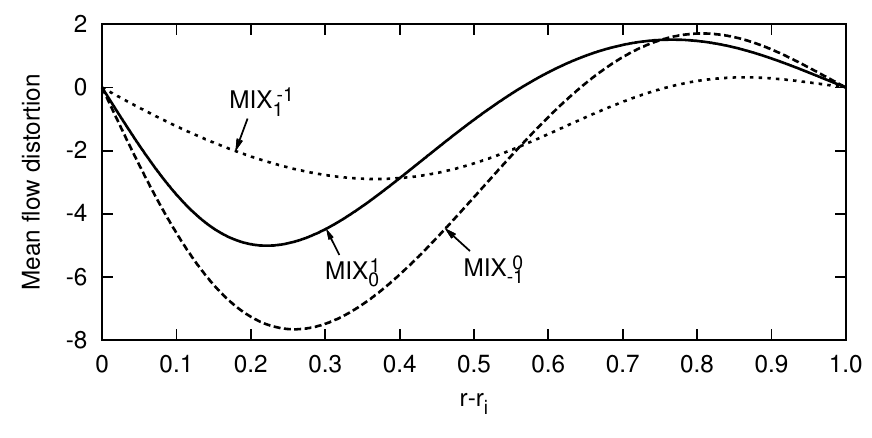}
\caption{Azimuthal mean flow distortion ($\overline{v}-v_b$) of the three different mixed
  modes shown in figure~\ref{fig:nc_bifdiag}b for $H=140$, and
  $(R_i,R_o)\approx(2074,1078)$.
}
\label{fig:azim_flow_dist}
\end{figure}

\begin{figure}
  \begin{center}
\begin{tabular}{cc}
  (a) & (b)\\
    \includegraphics[width=.45\linewidth,clip=]{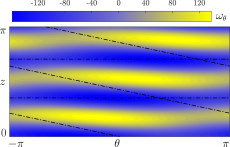}&
  \includegraphics[width=.45\linewidth,clip=]{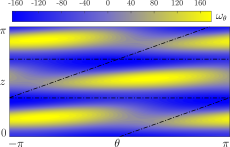}
\end{tabular}
\begin{tabular}{c}
  (c)\\
    \includegraphics[width=.45\linewidth,clip=]{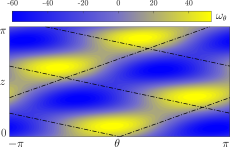} 
\end{tabular}
\end{center}
\caption{Colourmaps of the total azimuthal vorticity
  ($\omega_{\theta}=\partial_zu-\partial_rw$) distribution at the mid
  radial plane $r=r_i+0.5$ of the three mixed-mode {\sc mri}
  solutions.
  (a)
  $\textsc{mix}^{0}_{1}$, (b) $\textsc{mix}^{0}_{-1}$, and (c)
  $\textsc{mix}^{1}_{-1}$.}
\label{fig:mix_anticyclonic}
\end{figure}

All three mixed-mode branches bifurcate supercritically. A very
remarkable feature of these mixed modes is that some of them have
larger torque than the spirals. This aspect is of special relevance to
the study of astrophysical accretion disks, as it is a paramount
requirement for the large outward angular momentum flux that is
believed to be the key to the observed rate of inward mass
accretion. The {\sc mix}$^0_{-1}$ mode possesses the largest torque of
all mixed modes, as is also clear from the azimuthal mean flow
distortion shown in figure~\ref{fig:azim_flow_dist} for $H=140$,
$(R_i,R_o)\approx(2074,1078)$.
Figure~\ref{fig:mix_anticyclonic} shows the corresponding total
azimuthal vorticity distribution of the three mixed modes, represented
through $\theta$--$z$ plane colourmaps at mid gap $r=r_i+0.5$.  As
expected, mode {\sc mix}$^0_{-1}$ has the strongest flow field
perturbation, clearly reflected in the colour bar range of the
panels. The visualisations shown in
figures~\ref{fig:mix_anticyclonic}a and \ref{fig:mix_anticyclonic}b
for {\sc mix}$^1_{0}$ and {\sc mix}$^0_{-1}$, respectively, are
reminiscent of wavy Taylor vortex flow (see e.g. \citet{ALS86}) except
that the patterns are tilted and wavy vortex pairs accumulate an
azimuthal phase shift as they pile up in the axial direction. The
reason for this is that one of its constituents is a zero-pitch
spiral, which corresponds to a toroidal-vortex-pair solution much like
Taylor vortices, while the superposition of a spiral mode generates
the tilted azimuthal modulation. Meanwhile, the structure of the {\sc
  mix}$^{-1}_{1}$ mode shown in figure~\ref{fig:mix_anticyclonic}c are
evocative of the wavy spiral solution found in the hydrodynamic
studies by \citet{AlHo10} and \citet{DeAlt2013}.

\section{From counter-rotation to the cyclonic super-rotation regime}
\label{sec_counter_super}
\begin{figure}
\centering
\includegraphics[scale=1.2]{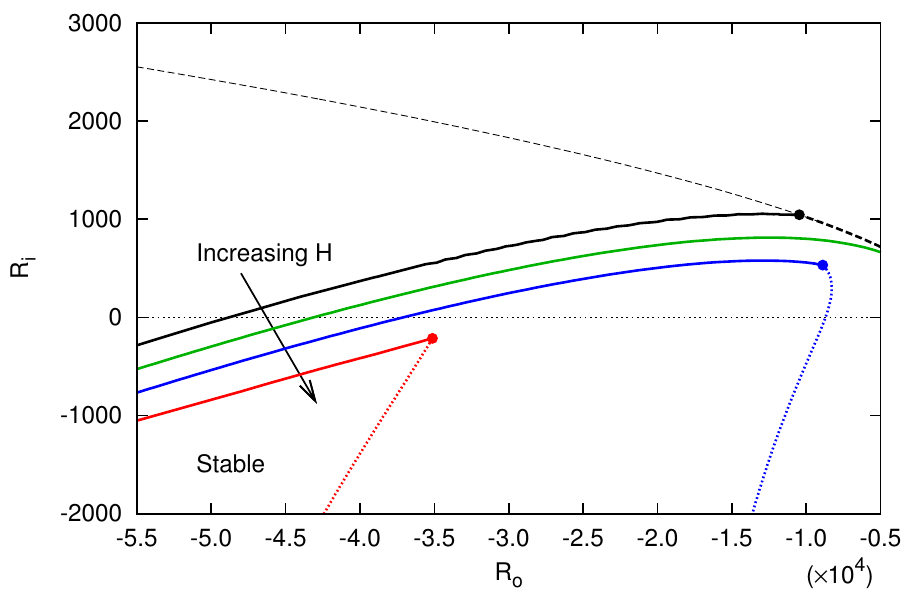} 
\caption{Neutral curves for $H=0, 40, 60, 84$ (black, green, blue,
  red, respectively) and $\eta=0.1$. The wavenumber pairs $(k,m)$ are
  optimized to detect the most unstable eigenvalue.  Neutral curve for
  the classical non-axisymmetric hydrodynamic mode (dashed black) is shown
  alongside those for the {\sc d}17 mode ({solid black}) and the {\sc mri}
  mode (dotted). While $m$ is large and obeys the asymptotic result by
  \citet{Deguchi_JFM17} for the classical neutral curve, the {\sc d}17
  and {\sc mri} neutral curves have typically $m=\pm 1$. Bicritical
  points, where direct bifurcation of mixed-mode solution branches are
  expected, are indicated with filled circles.}
\label{fig:neutralsuperSpa}
\end{figure}

In this section we focus our attention on the bifurcations arising on
the left-half plane of figure\,\ref{fig:param_plane}, as some of the
instabilities carry on to the super-rotation regime.  For $\eta=0.1$,
figure\,\ref{fig:neutralsuperSpa} outlines the neutral curves obtained
from linear stability analyses corresponding to different levels of
magnetization. We begin our analysis by first focusing on the purely
hydrodynamic case in the absence of magnetic effects. Shown in
figure~\ref{fig:neutralsuperSpa} are the classical neutral curve
(dashed black) alongside the neutral curve for the {\sc d}17 mode
(solid black), recently dicovered by \citet{Deguchi_PRE17} through
linear stability analysis.

Along the classical boundary, the critical value of $R_i$ increases
with $\lvert R_o\rvert$, which is consistent with extensive numerical
evidence as well as physical insights and the large Reynolds number
formal asymptotic result \citep{EsGr96,GroLoSu16,Deguchi_JFM17}. As a
consequence, the neutral curve, which corresponds to a
non-axisymmetric leading mode with large $m$ (a spiral), cannot be
continued across the line $R_i=0$ into the super-rotation regime.

In contrast, the neutral curve for the {\sc d}17 mode, typically with
$m=\pm1$, does indeed extend to the cyclonic super-rotation
regime. The reason for choosing such a low value of the radius ratio
($\eta=0.1$) follows from the observation that the curve shifts to
very high counter-rotation rates as $\eta$ is increased and narrower
gaps are considered. For instance, taking $R_i=0$ and $\eta=5/7$
pushes the critical $R_o$ value to $O(10^7)$, whereas for $\eta=0.1$
it remains within order $O(10^4)$. In
figure~\ref{fig:neutralsuperSpa}, the classical stability threshold
(dashed black) and the new one set by the neutral curve of the {\sc
  d}17 mode (solid black) meet at a bicritical point (black filled
circle) located at $(R_i,R_o)\approx(1045, -10434)$, with associated
critical wavenumbers $(m_1,k_1)=(1,40.6)$ and $(m_2,k_2)=(1,1.102)$,
respectively. The critical axial wavenumbers $k_1$ and $k_2$
associated to either mode differ significantly, which explains why the
latter, mode {\sc d}17, escaped detection for so long.  The asymptotic
theory provided by \citet{Deguchi_JFM17} formally proved that the
critical axial wavenumber of the classical mode gets asymptotically
large for increasing Reynolds numbers, while that for the {\sc d}17
mode seems to be insensitive to Reynolds number variations.

\begin{figure}
\begin{center}
\begin{tabular}{cc}
%  \raisebox{0.41\linewidth}{(a)} & \includegraphics[width=0.65\linewidth,clip=]{fig7a.eps} \\
%  \raisebox{0.41\linewidth}{(b)} & \includegraphics[width=0.65\linewidth,clip=]{fig7b.eps}
  \raisebox{0.41\linewidth}{(a)} & \includegraphics[width=0.65\linewidth,clip=]{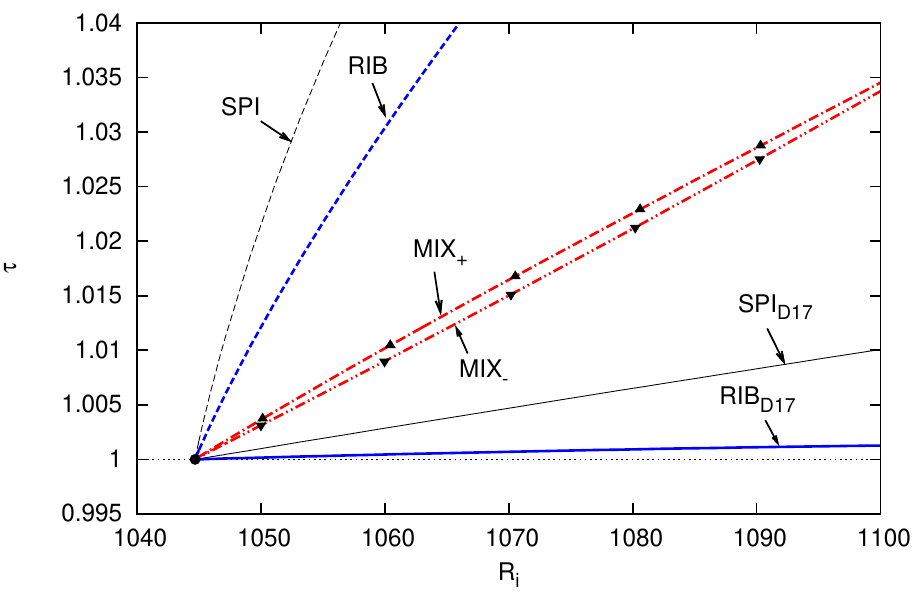} \\
  \raisebox{0.41\linewidth}{(b)} & \includegraphics[width=0.65\linewidth,clip=]{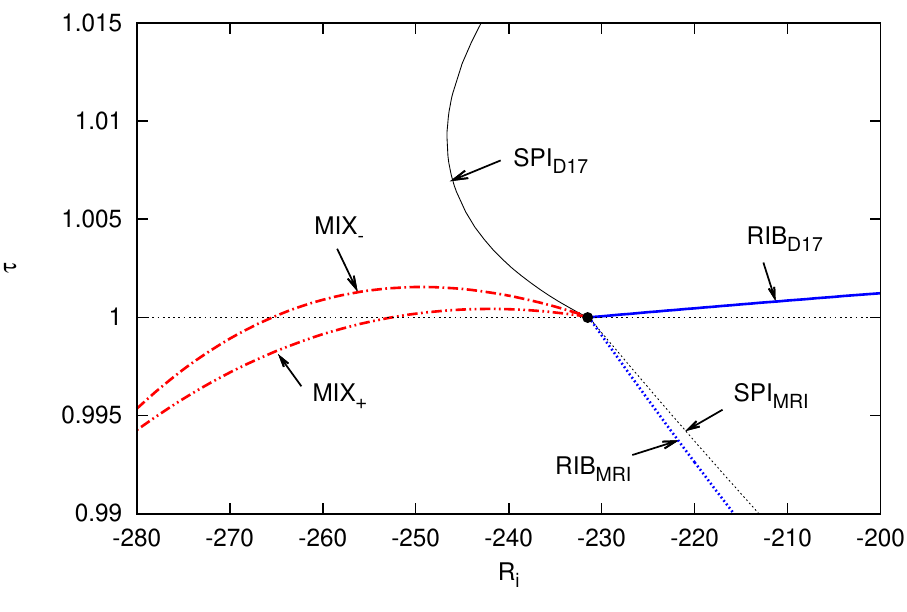} 
\end{tabular}
\end{center}
\caption{Bifurcation diagrams of spirals, ribbons and mixed nonlinear
  modes in super- and counter-rotation configurations. (a) Purely
  hydrodynamic case, in counter-rotation with $R_o=-10434$ and $H=0$.
  The black circle indicates the linear bicritical point at
  $R_i=1045$.  A bunch of mixed-mode
  solutions computed with the alternative Petrov-Galerkin code are
  marked with triangles. (b) Magnetised case in the super-rotation regime with
  $R_o=-35150$ and $H=84$.  The bicritical point at $R_i=-231.5$ is
  indicated with a filled black circle.
}\label{fig:SuperbifSPa}
\end{figure}

The various nonlinear solution branches that bifurcate from the
bicritical point (black filled circle in
figure~\ref{fig:neutralsuperSpa}) are shown in
figure~\ref{fig:SuperbifSPa}a. All branches bifurcate supercritically.
The black curves correspond to the spiral solutions
for the classical mode ({\sc spi}, dashed) and the {\sc d}17 mode
({\sc spi}$_{\text{D17}}$, solid). The structure of the single-mode
solutions, the nonlinear spirals, are qualitatively identical to the
linear neutral mode \citep*[see][]{Deguchi_PRE17}.  Since the system
is symmetric with respect to axial reflections ($z \rightarrow -z$),
spiral modes become neutral in pairs, with exact opposite pitch.  As a
result, there are actually four modes that simultaneously become
neutral at the linear bicritical point (the black filled circle in
figure~\ref{fig:SuperbifSPa}a). Consequently, there exist six mixed
modes arising from all possible combinations of modes taken in pairs.
Two of them merely correspond to ribbon solutions (blue
curves in figure~\ref{fig:SuperbifSPa}a), labelled
as {\sc rib} (dashed, {\sc spi}-{\sc spi} interaction) and {\sc
  rib}$_{\text{D17}}$ (solid, {\sc spi}$_{\text{D17}}$-{\sc
  spi}$_{\text{D17}}$ interaction), and listed in
table~\ref{tab_samples}. While ribbon solutions can be computed in a
rectangular domain, all other mixed modes require the use of the
annular-parallelogram domain. In fact, only two of the four remaining
modes actually require computation, as the other two can be easily
obtained from simple $z$-reflection and, since the torque is invariant
under this symmetry operation, the solution branches are exactly
coincident. The branches corresponding to these mixed-mode solutions
are painted in red in figure~\ref{fig:SuperbifSPa}a.  The branch
labelled as {\sc mix}$_+$ originates from the interaction of two modes
with pitches of the same sign, while the one labelled {\sc mix}$_-$
arises from the nonlinear coupling of modes with opposite sign, as
reported in table~\ref{tab_samples}.

\begin{figure}
\centering
\begin{tabular}{cc}
%\raisebox{0.3\linewidth}{(a)} & \includegraphics[scale=1.2]{fig8a.eps}\\
%\raisebox{0.3\linewidth}{(b)} & \includegraphics[scale=1.2]{fig8b.eps} 
\raisebox{0.3\linewidth}{(a)} & \includegraphics[scale=1.2]{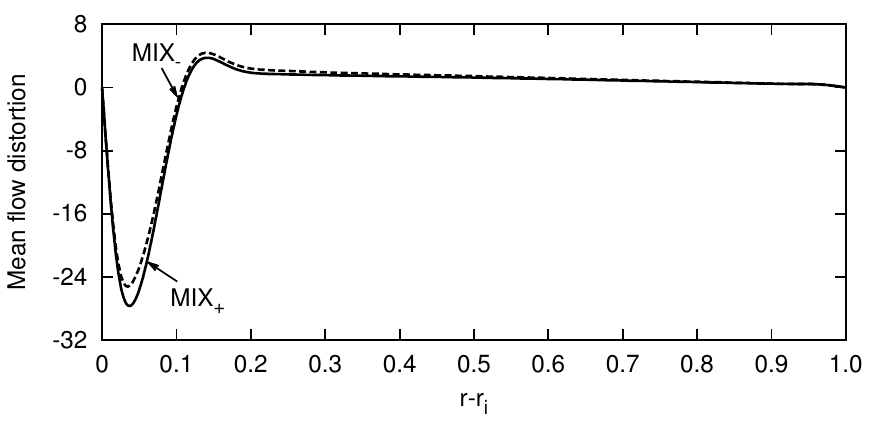}\\
\raisebox{0.3\linewidth}{(b)} & \includegraphics[scale=1.2]{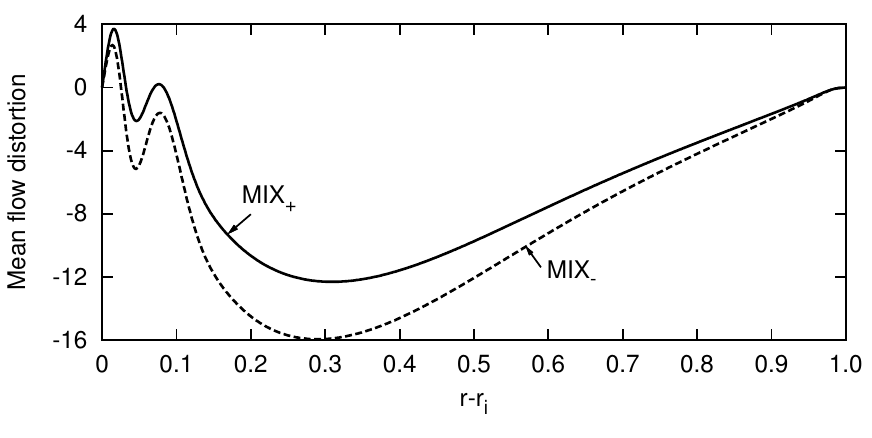} 
\end{tabular}
\caption{Azimuthal mean flow distortion ($\overline{v}-v_b$) of mixed-mode solutions. (a) Purely hydrodynamic case
  at $R_i=1100$ from
  figure~\ref{fig:SuperbifSPa}a. (b) Magnetised case at
  $R_i=-280$ and $H=84$ from figure~\ref{fig:SuperbifSPa}b. 
}
\label{fig:mean23Spa}
\end{figure}

\begin{figure}
\begin{center}
\begin{tabular}{cc}
  (a) & (b)\\
\includegraphics[width=.45\linewidth,clip=]{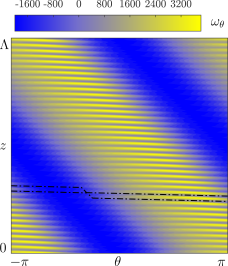}& 
\includegraphics[width=.45\linewidth,clip=]{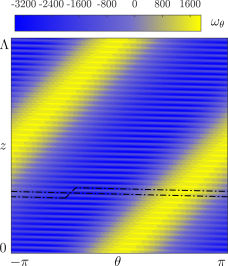} 
\end{tabular}
\end{center}
\caption{Visualisation of the total azimuthal vorticity
  ($\omega_{\theta}=\partial_zu-\partial_rw$) at $r=r_i+0.05$ for the
  purely hydrodynamic mixed mode solutions at $R_i=1100$ in
  Fig.\,\ref{fig:SuperbifSPa}(a). $\Lambda=2\pi/1.002 \approx 6.27$.
  (a) {\sc mix}$_-$ and (b) {\sc mix}$_{+}$. The
    corresponding mean flow distortion was shown in
    Fig.\,\ref{fig:mean23Spa}(a).}
\label{fig:VISyzOMIX}
\end{figure}

All hydrodynamic results reported in this work and initially computed
with a code based on the hydromagnetic-potential formulation
(\ref{potentials}) have been reproduced using the independently
developed solenoidal Petrov-Galerkin parallelogram formulation
(\ref{specapx}).  A few nonlinear solutions at
selected values of the parameters have been chosen and indicated with
triangles in figure~\ref{fig:SuperbifSPa}a to convey the excellent
qualitative agreement between the two methods employed for the
computations. Quantitative comparison shows that the torque
discrepancy stays below $0.06\%$ for all the purely hydrodynamic
nonlinear mixed-mode solutions computed.  Figure~\ref{fig:mean23Spa}a
represents the azimuthal mean flow distortion for the mixed modes at
$(R_i,R_o)=(1100,-10434)$.  The distortion is the most significant in
the vicinity of the inner cylinder, which indicates that the
perturbation is strongest in this region.  For the same values of the
parameters, figure~\ref{fig:VISyzOMIX} shows azimuthal vorticity
colourmaps for both mixed modes on an unwrapped radial plane at
$r=r_i+0.05$.
The observed flow structure is very different from any of the
mixed-mode solutions reported by \citet{DeAlt2013}.  The observed
small-large scale interaction reminds of the stripe pattern that is
characteristic of intermittent spiral turbulence
\citep{MeMeAvMa09_A,Dong2009}. While the Reynolds numbers and the gap
are too large to claim there exists any relation between the mixed
modes presented here and spiral turbulence, the similarity of the
patterns indicates that spiral turbulence might indeed be supported by
mixed-mode solutions of very different pitches as the ones
investigated here in a completely different setting.

Now we turn our attention to the magnetised problem, where we will
impose an external magnetic field with a strictly azimuthal
orientation.  For the {\sc mri} studies in an annulus, the azimuthal
base magnetic field is typically the weighted superposition of
$r^{-1}$ and $r$ components. The respective coefficients can be tuned
by an appropriate uniform current imposed within the inner and outer
cylinders.  \citet{RSGS16,RSGS18} considered two extreme cases:
$B_b(r)\propto r^{-1}$ (i.e. there is no
  current between the cylinders) and $B_b(r)\propto r$ (i.e. the axial current is homogeneous between the cylinders).  The
latter also receives the alternative name $z$-pinch, and is known to
become unstable for sufficiently large Hartmann number even with both
cylinders at rest \citep{Tayler57}. As the \textit{Tayler instability}
does not exist for the current-free case, the behaviour of the neutral
curve for small Reynolds numbers must necessarily
be quite different from that for the homogeneous-current case.
\citet{RSGS16,RSGS18} found that for $\eta\gtrsim 0.8$, the neutral curves behave qualitatively alike in both
  cases when Reynolds numbers are moderately large, thereby
suggesting that super-{\sc amri} is rather insensitive to the choice
of the azimuthal magnetic field profile.

We have confirmed that the {\sc d}17 mode is stabilised by both the
current-free and the $z$-pinch cases. Nonetheless,
  when the two azimuthal magnetic field components exist
  simultaneously, the {\sc d}17 mode can be destabilised, as clearly
  illustrated by the behaviour of the neutral stability curves in
  figure~\ref{fig:neutralsuperSpa}. Here the specific magnetic field
  profile used is
\begin{eqnarray}
B_b(r)=\frac{r_i}{r}-\frac{r}{r_o},\qquad C_b(r)=0.
\label{baseflowkengo}
\end{eqnarray}
Although the arguments by \citet{RSGS16,RSGS18} for
  the current-free case may not be applicable to the large gap
  $\eta=0.1$ we tackle here, a MRI does indeed arise when a current is
  considered. This phenomenon had already been anticipated by a
  locally periodic approach \citep[see][]{Liu2006,Kirillov2014}, but
  the nature of the method used renders the approximation rather crude
  in view of the not-so-large critical wavenumbers we encounter
  here. As the Hartmann number is increased, the
super-{\sc amri} mode eventually takes over the classical mode, and
hence changes the character of the bicritical point.  In view of
figure~\ref{fig:neutralsuperSpa} at $H=60$, the double critical point
is already the result of the interaction of the {\sc d}17 and the
super-{\sc amri} modes.  The base flow remains stable within the
region bounded by their respective stability thresholds.  Along the
combined neutral curve, the critical axial wavenumber experiences a
discontinuous leap across the bicritical point, whence it must be
inferred that the two instability mechanisms are indeed distinct.
By further increasing the Hartmann
number, the bicritical point moves towards and eventually crosses into
the super-rotation regime. We have determined that the bicritical
point crosses the $R_i=0$ line somewhere between $H=80$ and $H=84$.

The nonlinear solution branches issued from the bicritical point
$(R_i,R_o)\approx (-231.5, -35150)$ at $H=84$ have been computed in
the same way they were for the strictly hydrodynamic case studied
above.  The critical wavenumbers of the magnetised {\sc d}17 mode at
this point are $(m,k)=(1,0.616)$, while those for the super-{\sc amri}
modes are $(m,k)=(1,1.984)$, which entails flow structures of similar
sizes.  The bifurcation diagram of figure~\ref{fig:SuperbifSPa}b has
been obtained by varying $R_i$ at constant $R_o$.  To the {right
  (left)} of the linear critical point, the base flow is unstable to
the {\sc d}17 (super-{\sc amri}) mode.  The imposed azimuthal magnetic
field does not break any of the inherent symmetries of the
hydrodynamic Taylor-Couette system so that, as in the hydrodynamic
case discussed above, there still arise two spirals (along with their
mirror-images), two ribbons, and two mixed modes (and mirror
images). See table~\ref{tab_samples} for an account of all modes. Both
nonlinear spiral branches (black curves: solid for {\sc
  spi}$_{\text{D17}}$, dashed for {\sc spi}$_{\text{MRI}}$) bifurcate
\textit{subcritically}, in the sense that they exist when the
corresponding linear mode is stable. However, this is only true while
their amplitude remains small, and the branch associated with the {\sc
  d}17 mode turns back in a saddle-node bifurcation towards lower
$|R_i|$. The {\sc rib}$_{\text{D17}}$ (solid blue) and {\sc
  rib}$_{\text{MRI}}$ (dashed blue) solution branches exist both to
the right of the critical point (blue curves in
figure~\ref{fig:SuperbifSPa}b). We note in passing that the super-{\sc
  amri}-type ribbon solutions found by \citet{RSGS16} were shown
stable by direct numerical simulation.  The mixed-mode solution
branches (red curves) come in two types, namely {\sc mix}$_+$ 
and {\sc mix}$_-$, depending on whether the interacting modes
have the same or opposite pitch, respectively. Unlike all other
solution branches issued from the bicritical point, these extend to
large $|R_i|$, and may thus govern the dynamics within the region of
the super-rotation regime closest to solid body rotation.

As clear from figure~\ref{fig:SuperbifSPa}b, the nonlinear solution
branches associated with the super-{\sc amri} instability have the
unforseen property that the torque is reduced with respect to the
laminar base value.  The dependence of torque on Reynolds number
associated with the two mixed modes follows very similar trends. The
torque initially grows away from the bifurcation point as the branches
dive deep into the super-rotation regime, but the trend is soon
reversed and the torque eventually drops below laminar values.

The reason for torque reduction can
  be understood from the energy balance, since torque corresponds to
  one of the energy input mechanisms.  The perturbation energy budget
  can be found by integrating
  $\widetilde{\mathbf{v}}\cdot$(\ref{limmo})+$H^2\widetilde{\mathbf{b}}\cdot$(\ref{limind}).
  For travelling-wave-type solutions perturbation energy must be
  time-independent and thus the balance
\begin{eqnarray}
\left \langle r \left (\frac{v_b}{r} \right
)'\widetilde{u}\widetilde{v} \right \rangle -H^2\left \langle
r^{-1}(rB_b)'(\widetilde{a}\widetilde{v}-\widetilde{u}\widetilde{b})\right
\rangle=\langle \widetilde{\mathbf{v}}\cdot \nabla^2
\widetilde{\mathbf{v}}\rangle+H^2\langle \widetilde{\mathbf{b}}\cdot
\nabla^2 \widetilde{\mathbf{b}} \rangle \label{eng1}
\end{eqnarray}
should be satisfied.  Here the angle brackets denote integration over
the annular-parallelogram domain.  The first term in the right and
left hand sides are related to the torque, since integration of
$v_b\mathbf{e}_{\theta}\cdot$(\ref{limmo}) yields
\begin{eqnarray}
-\left \langle r \left (\frac{v_b}{r} \right )'\widetilde{u}\widetilde{v} \right \rangle
=\langle v_b \mathbf{e}_{\theta}\cdot \nabla^2 \widetilde{\mathbf{v}}\rangle
\end{eqnarray}
and integration by parts of the first term in the right hand side
results in
\begin{eqnarray}
\langle \mathbf{v}\cdot \nabla^2 \mathbf{v}\rangle=\left
(r_o^{-1}R_o-r_i^{-1}R_i \right )T-\langle |\nabla \mathbf{v}|^2
\rangle.\label{torqueeq}
\end{eqnarray}
The energy balance equation (\ref{eng1}) then becomes
\begin{eqnarray}
-H^2\langle \widetilde{\mathbf{b}}\cdot \nabla^2 \widetilde{\mathbf{b}} \rangle
-H^2\left \langle r^{-1}(rB_b)'(\widetilde{a}\widetilde{v}-\widetilde{u}\widetilde{b})\right \rangle=
\left (r_o^{-1}R_o-r_i^{-1}R_i \right )T-\langle |\nabla \mathbf{v}|^2 \rangle.\label{eng2}
\end{eqnarray}

Showing that torque cannot decrease below the laminar
  value for purely hydrodynamic Taylor-Couette flow is a
  straightforward exercise, because the terms on the left hand side
  are identically zero.  The calculus of variations can then be
  used to prove that the minimum value of the functional
  $\mathbf{F}(\mathbf{v})=\langle |\nabla \mathbf{v}|^2 \rangle$ under
  the divergence-free constraint for $\mathbf{v}$ is realised by the
  solution to the Stokes equation \citep[see][for example]{DoGi1995},
  namely the laminar Couette solution. Imposing $\langle |\nabla
  \mathbf{v}|^2 \rangle \geq \langle |\nabla \mathbf{v}_b|^2
  \rangle=\left (r_o^{-1}R_o-r_i^{-1}R_i \right )T_b $ on
  (\ref{torqueeq}) demands that $\tau \geq 1$ for the purely
  hydrodynamic case.

Moreover, the balance equation further leads to the
  conclusion that torque reduction cannot occur at all if the base
  magnetic field is current free, as this entails that the second term in the left hand side of
  (\ref{eng2}) is absent.  The proof is again straightforward as
  integration by parts shows that $-H^2\langle
  \widetilde{\mathbf{b}}\cdot \nabla^2 \widetilde{\mathbf{b}} \rangle$
  is positive definite.  The resulting inequality
  \begin{equation}
    \begin{array}{rcl}
      \left (r_o^{-1}R_o-r_i^{-1}R_i \right )T & \geq & \left
      (r_o^{-1}R_o-r_i^{-1}R_i \right )T+H^2\langle
      \widetilde{\mathbf{b}}\cdot \nabla^2 \widetilde{\mathbf{b}}
      \rangle =\langle |\nabla \mathbf{v}|^2 \rangle\\ & \geq &
      \left (r_o^{-1}R_o-r_i^{-1}R_i \right )T_b,
      \end{array}
\end{equation}
yields again $\tau \geq 1$. This outlines the necessity of a base
current field if torque reduction is to be observed.

\begin{figure}
  \begin{center}
\begin{tabular}{cc}
  (a) & (b)\\
    \includegraphics[width=.40\linewidth,clip=]{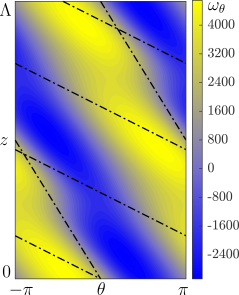}&
  \includegraphics[width=.40\linewidth,clip=]{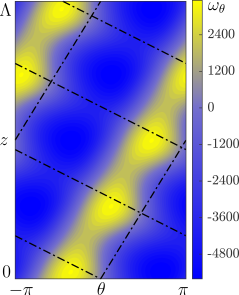}
\end{tabular}
\end{center}
\caption{Visualisation of the total azimuthal vorticity
  ($\omega_{\theta}=\partial_zu-\partial_rw$) at $r=r_i+0.05$ for the
  magnetized mixed mode solutions at $R_i=-280$ in
  Fig.\,\ref{fig:SuperbifSPa}(b). $\Lambda=2\pi/0.616\approx 10.2$ (a)
  {\sc mix}$_{-}$ and (b) {\sc mix}$_{+}$. The corresponding mean flow
  distortion was shown in Fig.\,\ref{fig:mean23Spa}(b).}
\label{fig:VISyzOAB}
\end{figure}

As seen in figure~\ref{fig:SuperbifSPa}b, the torque reduction is
stronger for the {\sc mix}$_+$ mode, reflecting the fact that the
perturbation is slightly larger for that mode.  This is evidenced by
figure~\ref{fig:mean23Spa}b, where the azimuthal mean flow distortion
across the gap is plotted at $R_i=-280$.  The oscillatory modulation
of the flow near the inner cylinder is responsible for the torque
reduction and is driven by the vortex structure near the inner
cylinder, as shown in the $\theta$-$z$ sections of
figure~\ref{fig:VISyzOAB}. Here again we choose $r=r_i+0.05$, very
close to the inner cylinder, as the reference radius.  As expected,
the perturbation of the {\sc mix}$_+$ mode has larger amplitude than
that for the {\sc mix}$_-$ mode. The flow patterns are similar to
those of the wavy spiral computed by \citet{AlHo10} and
\citet{DeAlt2013}, because the critical wavenumbers of the interacting
modes are of comparable size.

\section{Conclusions}
\label{sec_conclusions}
We have investigated nonlinear mode competitions in the {\sc mhd}
Taylor-Couette flow subject to predominantly azimuthal magnetic
fields.  For this purpose, a Newton solver devised by
\citet{DeAlt2013} for the Navier-Stokes equations in
annular-parallelogram domains has been extended for its application to
the inductionless limit of the {\sc mhd} equations.

For the anti-cyclonic regime (see figure\,\ref{fig:param_plane}), a
suitably adjusted weak axial magnetic field in addition to the
azimuthal field stimulates linear instability modes with $m=-1,0,1$,
as anticipated by \citet{HTR10}. Consistent with their results, we
find particularly rich nonlinear dynamics for $\delta\simeq0.04$.  In
section \S\ref{sec_anticyclonic}, we identified that there is a triple
critical point involving all three modes for $\delta \approx 0.0413$.
We have tracked the three nonlinear mixed-mode solution branches that
bifurcate simultaneously at the triple critical point using the
purposely devised Newton solver and arclength continuation.  Some of
the mixed-mode solutions possess a larger angular momentum transport
(they require application of higher driving torque to keep the
cylinders rotating) than the single-mode solutions they result from.
This increased transport of angular momentum makes these mixed-mode
solutions an interesting target for future study, as they might be
relevant in astrophysical flows involving accretion disks. In
particular, a better understanding of what might their role be in the
nonlinear dynamics of such flows will require direct numerical
simulations and experiments such as those by {\sc promise}
\citep{SGGRSSH06,SGGRSH07,RHSGGR06}.

In \S\ref{sec_counter_super}, we have studied mode competitions
involving the {\sc d}17 mode.  In the purely hydrodynamic case, there
is a point where both the classical spiral mode and the {\sc d}17 mode
become neutral simultaneously.  This bicritical point lies within the
counter-rotation regime (see figure\,\ref{fig:param_plane}).  The
corresponding mixed-mode solutions consist of an interesting stripe
pattern where the small-scale classical spirals are modulated by the
larger-scale structure of the {\sc d}17 mode. All purely hydrodynamic
results presented here are in excellent agreement with analogous
computations done with an independently developed Petrov-Galerkin code
devised by \citet{EPJSTMAMM07} and presently extended to allow
computation of mixed-mode travelling-rotating wave solutions in
annular-parallelogram domains. This code is better suited for the
study of large-scale pattern formation in Taylor-Couette flow and
includes not only travelling-rotating-waves Netwon-Krylov matrix-free
solver (thus being capable of handling a much larger amount of degrees
of freedom), but also stability analysis, a solver for modulated
travelling-rotating waves and pseudo-arclength continuation of
solution branches adapted from \citet{MeMe2015}, and also direct
numerical simulation. Details of this second code and its adaptation
to annular-parallelogram domains will be presented in our future work
in the study of large-scale pattern formation in Taylor-Couette
flow. The intricacies of the method reach beyond the scope of the
present study.

The application of an external azimuthal magnetic field alters the
picture obtained in the purely hydrodynamic case completely. The
non-axisymmetric super-{\sc amri} mode found by
\citet{RSGS16,RuGeHoSchuSte18,RSGS18} appears at moderate Hartmann
numbers and eventually outweighs the classical mode for sufficiently
strong magnetic fields.  We clearly show that the mechanisms behind
the magnetised {\sc d}17 mode and the super-{\sc amri} mode are
distinct.  
Destabilisation of the {\sc d}17 mode occurs for a given
external magnetic field profile (\ref{baseflowkengo}). As a result, an
increase of the Hartmann number gradually shifts the bicritical point
at which both modes are simultaneously destabilised towards
the super-rotation regime. This fact renders this
mode interaction interesting from an astrophysical point of view.
Several nonlinear solution branches are issued from the bicritical
point in both $R_i$ directions at fixed $R_o$. While spirals and
ribbons return towards the counter-rotation regime, the mixed-mode
solution branches plunge deep into the super-rotation regime.  

The solutions computed in \S\ref{sec_counter_super} show how the
complex interplay between the nonlinear shear-Coriolis and the
magneto-rotational instabilities can sometimes lead to torques lower
than that of the base flow. This surprising result
is in sharp contrast with what is typically assumed in purely
hydrodynamic shear-flow studies, where nonlinearity is known to
invariably enhance angular momentum transport.  This torque reduction
occurs even for finite $P_m$ and sub-rotation of the cylinders; see
Appendix \ref{appdrag}. In view of this result the torque reduction might be a
generic property of {\sc mhd} flows in the presence of shear and
Coriolis forces.  A particularly interesting potential application of
this phenomenon would be to design control strategies to reduce drag
on the curved boundary layers by imposing suitable magnetic fields.

We have analysed the {\sc d}17 / super-{\sc amri}
  mode interaction for a large gap $\eta=0.1$. Interesting as it
  would be, we have not attempted here to track these modes
  to smaller annulus gaps in the order $\eta\sim 0.8$ and test the
  robustness of the coalescence point of both instabilities. The
  extremely large Reynolds numbers at which the {\sc d}17 mode
  bifurcates in the narrow annulus case renders the task overly
  demanding from a computational point of view, if not altogether
  unaffordable.

Imposing more intense helical magnetic fields might also be an appealing topic for future research.
The two axisymmetric super-{\sc hmri} modes found recently by \citet{MaStHoRu19}, in combination with some of the modes studied here, may also yield rich interaction patterns worth exploring. While their type 2 super-{\sc hmri} mode belongs to the class of {\sc mri} requiring induction together with {\sc smri}, the type 1 super-{\sc hmri} mode is inductionless and might therefore coexist with the super-{\sc amri} mode and interact nonlinearly. Both types of super-{\sc hmri} modes might of course interact with the {\sc d}17 mode. 
For {\sc hmri}, the further consideration of current in the fluid brings about even richer instability phenomena as anticipated by a locally periodic approach \citep{Liu2006,Kirillov2013,Kirillov2014}. The interaction of short wavelength modes found using the locally periodic approach with longer wavelength modes such as {\sc d}17 or super-{\sc amri} modes would generate band-like patterns much as those in figure~\ref{fig:VISyzOMIX}.

Declaration of Interests. The authors report no conflict of interest.

\section{Acknowledgements}
\label{acknow}
KD's research was supported by Australian Research Council Discovery
Early Career Researcher Award DE170100171. RA, FM and AM research was
supported by the Spanish MINECO Grants FIS2016-77849-R,
FIS2017-85794-P, and the Generalitat de Catalunya
grant 2017-SGR-785. We greatfully acknowledge insightful comments and suggestions made by Prof. Rainer Hollerbach during the review process.

\appendix
\section{Basis function for the magnetic potentials}
Following Roberts (1964), we first determine the magnetic field for
the outer zones $r<r_i$ and $r>r_o$.  Within the perfectly insulating
walls, the magnetic field must have a potential $\varphi$ because
there is no current:
\begin{eqnarray}
\widetilde{a}=\varphi_r,\qquad
\widetilde{b}=r^{-1}\varphi_{\theta},\qquad
\widetilde{c}=\varphi_z.\label{magcont}
\end{eqnarray}
Since the magnetic field is solenoidal, the outer potential must
satisfy Laplace's equation.  Using the expansion
\begin{eqnarray}
\varphi=\sum_{n_1,n_2}\widehat{\varphi}_{n_1n_2}(r)\rme^{\rmi(n_1\xi_1+n_2\xi_2)},
\end{eqnarray}
it is easy to find that the solution $\widehat{\varphi}_{n_1n_2}(r)$
can be written down using the modified Bessel functions of the first
and second kind, $I_{\nu}(x), K_{\nu}(x)$, both of which satisfy
${x^2f''+xf'-(x^2+\nu^2)f=0}$.  The requirement that the potential is
analytic at $r=0$ determines the solution for $r<r_i$ as
\begin{subequations}\label{outpot}
\begin{eqnarray}
\widehat{\varphi}_{n_1n_2}=
\left \{
\begin{array}{c}
I_{|A_{n_1n_2}|}(|B_{n_1n_2}| r) \qquad \text{if} \qquad
|B_{n_1n_2}|\neq 0,\\~\\
r^{|A_{n_1n_2}|}\qquad \text{if} \qquad |B_{n_1n_2}|= 0,
\end{array}
\right .
\end{eqnarray}
whilst if the amplitude of the potential decays for large $r$, the solution for $r>r_o$ is
\begin{eqnarray}
\widehat{\varphi}_{n_1n_2}=
\left \{
\begin{array}{c}
K_{|A_{n_1n_2}|}(|B_{n_1n_2}| r) \qquad \text{if} \qquad |B_{n_1n_2}|\neq 0,\\~\\
r^{-|A_{n_1n_2}|}\qquad \text{if} \qquad |B_{n_1n_2}|= 0.
\end{array}
\right .
\end{eqnarray}
\end{subequations}
Here we have used the shorthand notation ${A_{n_1n_2}=n_1m_1+n_2m_2}$
and ${B_{n_1n_2}=n_1k_1+n_2k_2}$. Note that the mean part
$\widehat{\varphi}_{00}$ must be zero, from the boundary conditions.

Across the cylinder walls, the magnetic field must be continuous.
Thus from (\ref{magcont}) and the outer potential solutions
(\ref{outpot}), the boundary conditions are found as
\begin{subequations}
\begin{eqnarray}
  \widehat{b}_{n_1n_2}=
  \frac{A_{n_1n_2}}{r_iB_{n_1n_2}}\widehat{c}_{n_1n_2}\qquad
  \widehat{a}_{n_1n_2}+
  \frac{\rmi \, Q_{n_1n_2}^-}{B_{n_1n_2}}\widehat{c}_{n_1n_2}=0,\qquad \text{at}~~~r=r_i,\\
\widehat{b}_{n_1n_2}=\frac{A_{n_1n_2}}{r_oB_{n_1n_2}}\widehat{c}_{n_1n_2}\qquad
\widehat{a}_{n_1n_2}+\frac{\rmi \, Q_{n_1n_2}^+}{B_{n_1n_2}}\widehat{c}_{n_1n_2}=0,\qquad \text{at}~~~r=r_o.
\end{eqnarray}
\end{subequations}
Here $Q_{n_1n_2}^{\pm}$ denotes the value of
$\partial_r\widehat{\varphi}_{n_1n_2}/\widehat{\varphi}_{n_1n_2}$ on
the walls:
\begin{subequations}
\begin{eqnarray}
Q_{n_1n_2}^-=\left \{
\begin{array}{c}
\displaystyle \frac{|A_{n_1n_2}|}{r_i}+\frac{|B_{n_1n_2}|
  I_{|A_{n_1n_2}|+1}(|B_{n_1n_2}|r_i)}{I_{|A_{n_1n_2}|}(|B_{n_1n_2}|r_i)},
\qquad \text{if} \qquad |B_{n_1n_2}|\neq 0,\\~\\
\displaystyle \frac{|A_{n_1n_2}|}{r_i},\qquad \text{if} \qquad |B_{n_1n_2}|=0,
\end{array} \right .\\
Q_{n_1n_2}^+=\left \{
\begin{array}{c}
\displaystyle \frac{|A_{n_1n_2}|}{r_o}-\frac{|B_{n_1n_2}|
  K_{|A_{n_1n_2}|+1}(|B_{n_1n_2}|r_o)}{K_{|A_{n_1n_2}|}(|B_{n_1n_2}|r_o)},\qquad
\text{if} \qquad |B_{n_1n_2}|\neq
0,\\~\\ \displaystyle -\frac{|A_{n_1n_2}|}{r_o},\qquad \text{if} \qquad
|B_{n_1n_2}|=0.
\end{array} \right .
\end{eqnarray}
\end{subequations}
After some algebra, we can find the boundary conditions for the
poloidal and toroidal potentials as
\begin{subequations}\label{bdfg}
\begin{eqnarray}
\widehat{f}_{n_1n_2}'+M^-_{n_1n_2}\widehat{f}_{n_1n_2}=0,\qquad
\widehat{g}_{n_1n_2}-\gamma_{n_1n_2} (r_i)
\widehat{f}_{n_1n_2}=0\qquad
\text{at}~~~r=r_i\\ \widehat{f}_{n_1n_2}'+M^+_{n_1n_2}\widehat{f}_{n_1n_2}=0,\qquad
\widehat{g}_{n_1n_2}-\gamma_{n_1n_2} (r_o)
\widehat{f}_{n_1n_2}=0\qquad \text{at}~~~r=r_o,
\end{eqnarray}
\end{subequations}
where
\begin{subequations}
\begin{eqnarray}
\gamma_{n_1n_2}
(r)=\frac{2A_{n_1n_2}B_{n_1n_2}}{B_{n_1n_2}^2r^2+A_{n_1n_2}^2},\\ M^-_{n_1n_2}=
\frac{r_i^2B_{n_1n_2}^2-A_{n_1n_2}^2}{r_i^2B^2_{n_1n_2}+A^2_{n_1n_2}}-
\frac{r_i^2B_{n_1n_2}^2+A_{n_1n_2}^2}{r_i^2Q_{n_1n_2}^-},\\
M^+_{n_1n_2}=\frac{r_o^2B_{n_1n_2}^2-A_{n_1n_2}^2}{r_o^2B_{n_1n_2}^2+A_{n_1n_2}^2}-
\frac{r_o^2B_{n_1n_2}^2+A_{n_1n_2}^2}{r_o^2Q_{n_1n_2}^+}.
\end{eqnarray}
\end{subequations}
The second boundary conditions in (\ref{bdfg}) suggest that the
functions $\widehat{g}_{n_1n_2}-\gamma_{n_1n_2}
\widehat{f}_{n_1n_2}$ must vanish on the walls, and thus we can use
$(1-y^2)T_l(y)$ to expand them.  The function $\widehat{f}_{n_1n_2}$
satisfies Robin's conditions on the walls as seen in the first
boundary conditions in (\ref{bdfg}). As shown in
\citet{Deguchi_JFM19a}, we can use the following
modified basis functions
\begin{eqnarray}
(1-y^2)T_l(y)+\alpha_{ln_1n_2}+\beta_{ln_1n_2} y,
\end{eqnarray}
where
\begin{subequations}
\begin{eqnarray}
\alpha_{ln_1n_2}=2\frac{(-1)^l(1+M^+_{n_1n_2})+(1-M^-_{n_1n_2})}{(1-M^-_{n_1n_2})M^+_{n_1n_2}
  -(1+M^+_{n_1n_2})M^-_{n_1n_2}},\\ \beta_{ln_1n_2}=-2\frac{(-1)^l
  M^+_{n_1n_2} +M^-_{n_1n_2}}{(1-M^-_{n_1n_2})M^+_{n_1n_2}
  -(1+M^+_{n_1n_2})M^-_{n_1n_2}}.
\end{eqnarray}
\end{subequations}

\section{Drag reduction of the wavy vortex flow}\label{appdrag}
Here we show that the significant drag reduction observed in section 4
occurs even for the Rayleigh unstable sub-rotation regime (see
 figure\,\ref{fig:param_plane}).  Moreover, the Prandtl number is not necessarily small to
observe this phenomena; here we choose $P_m=1$. The base magnetic
field (\ref{baseflowkengo}) is used.

We employ the narrow-gap limit $\eta\rightarrow 1$ in order to use the
full {\sc mhd} Cartesian code developed in \citet{Deguchi_JFM19b}.
Now we write $x=r_m \theta$, $y=(r-r_m)$ using the mid gap $r_m$.
When $\eta$ is close to unity, noting $y/r_m \ll 1$, we have
approximations
\begin{eqnarray}
v_b(r)-\Omega r=-Ry+\cdots, \qquad HB_b(r)=-B_0Ry+\cdots,
\end{eqnarray}
while keeping $\Omega=r_m^{-1}v_b(r_m)$, $R=2r_m^{-2}R_p$ and
$B_0=H/R_p$ as $O((1-\eta)^0)$ constants. (Note that the definition of
$R$ differs by factor of 4 from that used in \citet{Deguchi_JFM19b}, because
in this paper the gap is 2.)  The limiting system is the rotating
plane Couette flow in the cartesian coordinate $(x,y,z)$ with the
rotation rate $\omega=2\Omega/R$. The Rayleigh unstable region is
$\omega \in [0,1]$. For $B_0\neq 0$, the flow is subjected to a linear
magnetic field pointing in the streamwise direction.

The first few bifurcation sequence of hydrodynamic rotating plane
Couette flow is widely acknowledged \citep[see][for
  example]{Nag86,DaSchSchPea14}. Near the Rayleigh line $\omega=1$,
the Taylor-vortex flow {bifurcates} with the well-known axial critical
wavenumber {$k=$}3.117 as depicted by the dashed curve in Fig. 11.
Further bifurcation of the green solid curve is due to the
three-dimensional secondary instability of the Taylor-vortex and
called the wavy-vortex flow.

The other solid curves in Fig. 11 (blue for $B_0=0.5$, red for
$B_0=1$) show that with increasing $B_0$ the shear associated with the
wavy vortex flow is eventually reduced even below the laminar value.
Here note that this reduction only occurs when the flow is dependent
on $x$ (i.e. azimuthal direction).  For example Taylor-vortex flow is $x$-independent, and thus
the shear is unchanged whatever the value of $B_0$ is. This is because
the cross-streamwise components of the magnetic field needed to modify
the mean flow remains zero under the influence of the streamwise
magnetic field.

\begin{figure}
\centering
\includegraphics[scale=1.1]{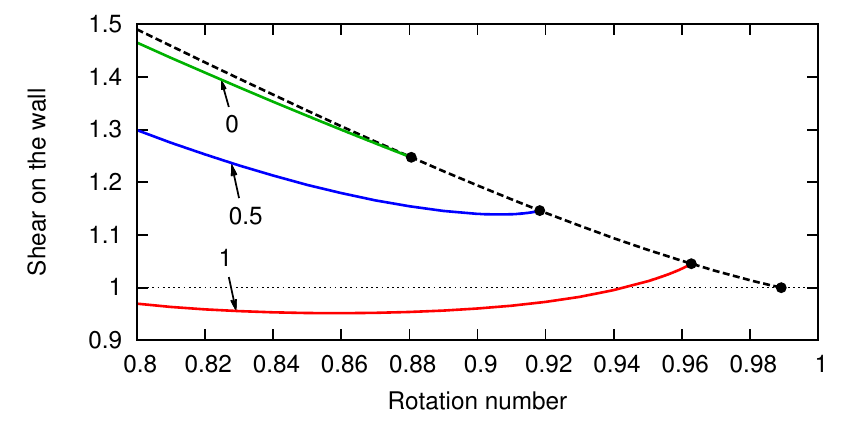} 
\caption{The narrow-gap computation for $P_m=1$, $R=400$, and the
  axial wavenumber {$k=3.117$}. Dashed curve: Taylor vortex
  flow. Solid curves: the wavy vortex flow
  branches. The streamwise wavenumber (i.\,e. $m/r_m$
    at the narrow-gap limit) is $2.2$. The values of $B_0$ are
  indicated by the arrows.  Horizontal axis is the rotation number
  $\omega$.  In the vertical coordinate, the shear on the wall is
  normalized by its laminar value (i.e. $\tau$ at the
    narrow-gap limit).}
\label{fig:longwave}
\end{figure}

\bibliographystyle{jfm}
\bibliography{local}

\end{document}